# Reconstruction of Potential Flight Paths for the January 2015 “Gimbal” UAP

Yannick Peings and Marik von Rennenkampff
*Independent Researchers*

**The “Gimbal” video is arguably the most recognizable publicly-available footage of unidentified anomalous phenomena (UAP). Recorded in January 2015 off the coast of Jacksonville, Florida, by a U.S. Navy F/A-18F Super Hornet’s AN/ASQ-228 ATFLIR targeting pod, the video shows an infrared-significant object skimming over clouds. Towards the end of the 34-second clip, the object appears to stop and rotate in mid-air. Naval aviators who participated in the event indicate that: (1) The UAP was within 10 nautical miles of the F/A-18F, (2) that, from the perspective of the aircrew’s top-down radar display, it was seen to stop and reverse direction with no radius of turn, and (3) that the UAP was accompanied by a formation of 4-6 other objects. Using data from the ATFLIR video, it is possible to reconstruct potential flight paths for the object as a function of distance. We show that, at the range provided by the aviators, potential flight paths align with eyewitness accounts: The object decelerates from a few hundred knots before rapidly reversing direction in a “vertical U-turn”. Such a maneuver would have been observed on the overhead radar display as an abrupt reversal of direction with no radius of turn. The highly anomalous flight path found at the range provided by the aircrew, along with the remarkable match between the reconstructed flight path, eyewitness recollections, and the object’s rotation, raises intriguing questions about the nature of the object. This is especially the case because, at this distance, no wings or infrared signatures consistent with conventional means of propulsion (e.g., an exhaust plume in the direction of flight) are visible. An alternative hypothesis, which proposes that Gimbal shows infrared “glare” from the exhaust of a conventional jet aircraft viewed approximately tail-on 30 nautical miles from the F/A-18F, is also discussed. According to this theory, the rotation observed in the video is an artifact of the ATFLIR targeting pod. Our goal is to provide an overview of analyses of the Gimbal encounter conducted by private citizens. We encourage aeronautics/aerospace experts to provide feedback so that a better understanding of the Gimbal UAP may be achieved.**

**Contact**
Yannick Peings, yannick.peings@gmail.com
Marik von Rennenkampff, Mvonrennenkampff@gmail.com

# I. Nomenclature

| | | |
|---|---|---|
| *ATFLIR* | = | Advanced Targeting Forward Looking Infrared |
| *Az* | = | Azimuth angle |
| *COMPTUEX* | = | Composite Training Unit Exercise |
| *DoD* | = | Department of Defense |
| *El* | = | Elevation angle |
| *FLIR* | = | Forward Looking Infrared |
| *FOV* | = | Field of View |
| *IAS* | = | Indicated Air Speed |
| *kt* | = | Knot (speed) |
| *LOS* | = | Lines of Sight |
| *Nm* | = | Nautical Mile |
| *ROT* | = | Rate of Turn |
| *SA* | = | Situational Awareness (radar display) |
| *TAS* | = | True Air Speed |
| *UAP* | = | Unidentified Anomalous Phenomena |
| *WSO* | = | Weapon Systems Officer |

# II. Introduction

On December 17, 2017, the *New York Times* reported[1] that the Department of Defense had been analyzing military reports of unidentified anomalous phenomena (UAP) since 2008. Accompanying the article were two videos recorded by a U.S. Navy F/A-18F Super Hornet in January 2015. One of the videos, entitled "Gimbal"[2] appears to show an infrared-significant object with no apparent control surfaces or means of propulsion skimming across clouds. The cockpit intercom (audio) recording indicates that the pilot and weapon systems officer (WSO) were surprised by what they observed, especially as the object appears to rotate approximately 90 degrees mid-flight. Of note, the incident coincided with naval aviators observing anomalous radar contacts (frequently corroborated via infrared sensors, and infrequently observed visually) on a daily basis in tightly-controlled training areas off the U.S. East Coast. Those contacts, observed over the course of years, exhibited anomalous flight characteristics, to include remaining stationary over a ground location in winds aloft or moving between 0.6 - 1.2 Mach for durations far exceeding[3] those of conventional fighter aircraft [1].

In the ensuing years, members of Congress took an interest in these (and other) incursions into sensitive airspace. Beyond drafting legislation establishing a new office to investigate UAP incidents, Senate Armed Services Committee professional staff members interviewed the weapon systems officer (WSO) who recorded the Gimbal encounter. The following summary of the Gimbal event is based on the WSO's briefing to congressional staff (Figure 1), as well as from the recollections of former U.S. Navy Strike Fighter Squadron (VFA) 11 F/A-18F pilot Lieutenant (LT) Ryan Graves, who was airborne during the Gimbal encounter.[4]

One evening in January 2015, multiple F/A-18Fs from VFA-11 participated in a large-scale air-to-air training mission approximately 300 miles off the coast of Jacksonville, Florida (see approximate location in Figure A1), as part of a pre-combat deployment Composite Training Unit Exercise (COMPTUEX) mission. Following the conclusion of the mission, as the aircraft returned one-by-one to the USS Theodore Roosevelt (CVN-71), one

---

[1] https://www.nytimes.com/2017/12/16/us/politics/pentagon-program-ufo-harry-reid.html
[2] https://www.navair.navy.mil/foia/sites/g/files/jejdrs566/files/2020-04/2%20-%20GIMBAL.wmv
[3] https://www.nytimes.com/2019/05/26/us/politics/ufo-sightings-navy-pilots.html
[4] https://thedebrief.org/devices-of-unknown-origin-part-ii-interlopers-over-the-atlantic-ryan-graves/

F/A-18F detected an air contact traveling west, in the direction of the Roosevelt (Figure 1, “an air contact [...] coming from the east”). The F/A-18F diverted from its return to the carrier to investigate the object.

According to the F/A-18F WSO’s briefing to congressional staff, the aircrew initially thought that the contact could be a land-based adversarial aircraft associated with the training exercise (Figure 1, “initially thinking [...] as part of the COMPTUEX scenario”). With a “stable trackfile” (i.e., radar lock), the aircrew determined that the contact was “not a ‘false hit’” (Figure 1), and subsequently acquired the object on the Super Hornet’s AN/ASQ-228 ATFLIR targeting pod, resulting in the 34-second Gimbal video.

Of note, the aircrew are heard commenting about a “fleet” of objects accompanying the Gimbal object and moving against a 120-knot (kt) wind. Former fighter pilot Graves, recalling the incident in subsequent interviews, stated that from the aviators’ situational awareness (SA) display (i.e., a top-down, god’s-eye view of nearby radar contacts fed by onboard and offboard sensors), the Gimbal object trailed a loose wedge formation of four to six objects that, due to their unique variations in target aspect, appeared on the radar display to resemble the anomalous objects observed daily by naval aviators off the U.S. East Coast.[4]

According to LT Graves, as viewed on the SA display, the “fleet” of objects then began a turn, ultimately conducting a 180-degree reversal in the direction of travel. The Gimbal object also reversed direction. This is corroborated in the WSO’s briefing to Congress (Figure 1, “He and his pilot detected an air contact [redacted] coming from the east and heading towards the ship,” which ultimately “headed back towards the east, and away from the [redacted/presumably ‘ship’]”). Unlike the “fleet,” however, the Gimbal object’s direction reversal occurred without a radius of turn. According to LT Graves, “the ‘Gimbal’ object that was following behind them suddenly stopped and waited for the wedge formation to pass. Then it tilted up like you can see in the clip, and that’s when my video cut out, but it just kept following the other five or six, doing like a racetrack pattern.”[4]

Of particular significance, aircrew involved/familiar with the encounter have high confidence that the Gimbal object was within 10 nautical miles (Nm) of the F/A-18F that recorded the video. This awareness of range to the target is hinted at in Figure 1 (“Closest point of intercept being approximately [redacted],” “Closest the air vehicles came to the [redacted]”) and comes from their direct observation of radar data. Through public and personal communication, LT Ryan Graves confirmed this range estimate provided by the WSO, refined to 6-8 Nm.[5] This is likely the range given in the incident report but, unfortunately, it is redacted in the available documentation (Figure 1, “Closest the air vehicles came to the [redacted] was approximately [redacted],” “the closest point of intercept being approximately [redacted]”).

With no official comment or explanation for the Gimbal video since its 2017 publication, members of the public have conducted detailed analyses of the footage, sparking vigorous debate about the nature of the object. On-screen data allows for reconstructions of potential flight paths for the object, and our objective here is to verify if paths that are consistent with what the aviators reported can be retrieved. This work is meant to present the status of research on this case in an objective manner, with the goal of spurring further research on this incident, and UAP more generally. Importantly, we encourage experts in aviation and military systems to provide their interpretations of the video. Section III presents the methodology that was followed to estimate potential flight paths for the object. The results and discussion of possible scenarios are discussed in Section IV.

[5] https://twitter.com/uncertainvector/status/1508144005871153155

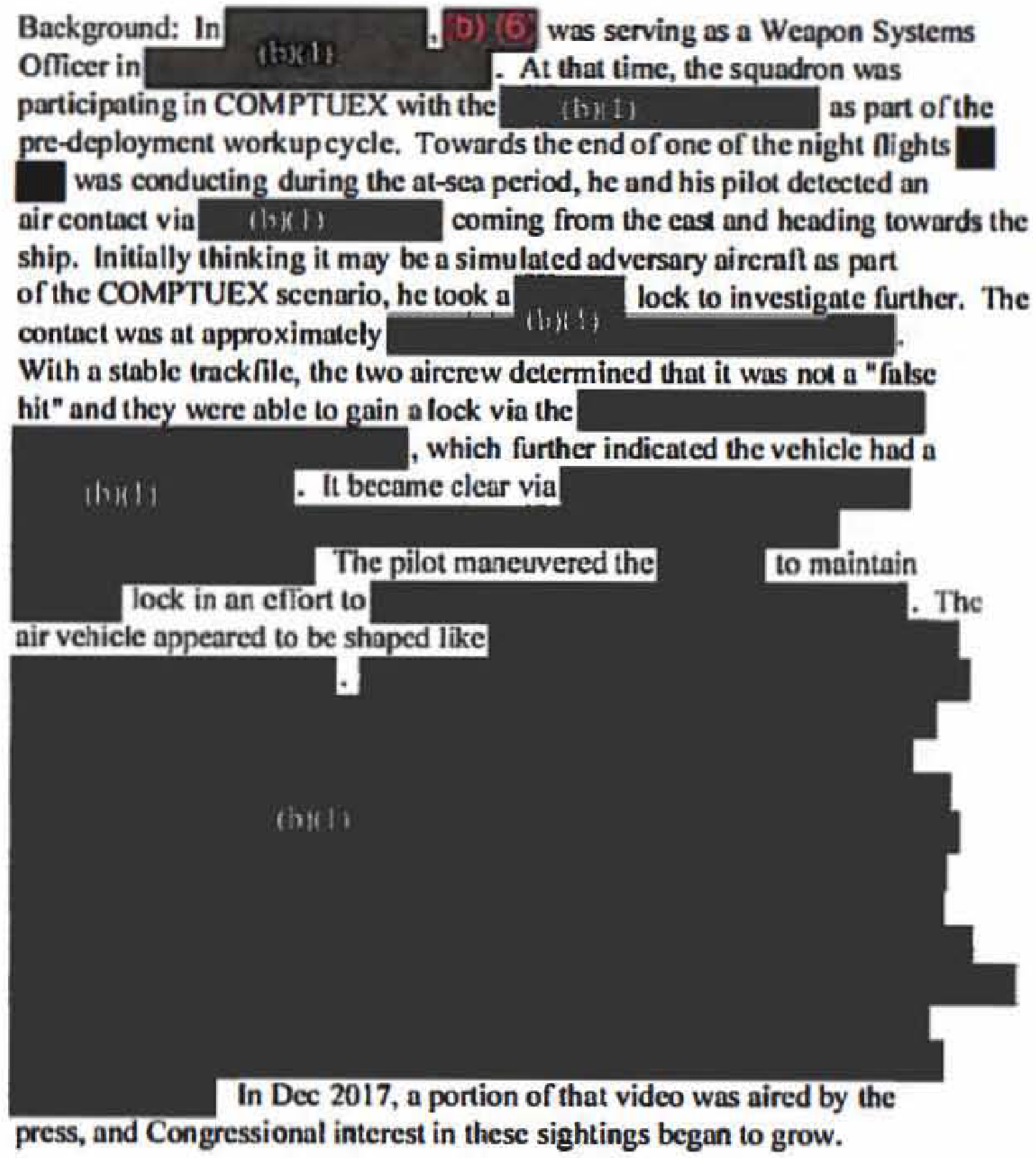

Background: In (b)(1), (b) (6) was serving as a Weapon Systems Officer in (b)(1). At that time, the squadron was participating in COMPTUEX with the (b)(1) as part of the pre-deployment workup cycle. Towards the end of one of the night flights was conducting during the at-sea period, he and his pilot detected an air contact via (b)(1) coming from the east and heading towards the ship. Initially thinking it may be a simulated adversary aircraft as part of the COMPTUEX scenario, he took a (b)(1) lock to investigate further. The contact was at approximately . With a stable trackfile, the two aircrew determined that it was not a "false hit" and they were able to gain a lock via the , which further indicated the vehicle had a (b)(1). It became clear via The pilot maneuvered the to maintain lock in an effort to . The air vehicle appeared to be shaped like .

(b)(1)

In Dec 2017, a portion of that video was aired by the press, and Congressional interest in these sightings began to grow.

~~SECRET//REL TO USA, FVEY~~

**2015-002 Cont:**

heading towards the (b)(1). Initially thinking it may be a (b) (1) as part of the (b) (1), he took a (b) (1) to investigate further. The contact was at approximately (b) (1). With a (b) (1), the two aircrew determined that it was (b) (1) and they were able to gain a (b) (1) (b) (1), which (b) (1) the vehicle had a (b) (1) (b) (1) It became clear via the (b) (1) that there were (b) (1) air vehicles flying (b) (1) (b) (1) type formation. The pilot maneuvered the (b) (1) to maintain (b) (1) in an effort to gather (b) (1) and try to make an ID. The air vehicle appeared to be shaped like a (b) (1), resembling some type of (b) (1). Maintaining (b) (1) on what appeared to be the (b)(1) craft, LT (b)(6) and his pilot noted what appeared to be very (b) (1) that it made. Because this event took place at night (b) (1) being used and the closest point of intercept being approximately (b) (1)(b) (1) with the craft(s) was never made. Instead, all of the aerial (b) (1) and headed back towards the east, away from the (b)(1). Closest the air vehicles came to the (b)(1) was approximately (b) (1). Once LT (b)(6) and his pilot were back on board the (b) (1) was viewed (b) (1) (b) (1) and (b) (1)), but nothing more was ever discussed or analyzed about the event after it occurred.

**Figure 1. Description[6,7] of the Gimbal incident by the weapon systems officer (WSO) who filmed the video. There are two available versions, each with different redactions.**

[6] https://documents2.theblackvault.com/documents/navy/DON-NAVY-2022-001613.pdf#page=6

[7] https://www.secnav.navy.mil/foia/readingroom/CaseFiles/UAP%20INFO/UAP%20DOCUMENTS/RF%20Reports%20Redacted%20(202301).pdf#page=26

## III. Methods

The ATFLIR video includes data that can be exploited to reconstruct potential flight paths for the Gimbal UAP. The critical missing data is the range (distance from the F/A-18F to the object), as well as the evolution of the range over the course of the 34-second video (i.e., is the F/A-18F getting closer to the object, further away, or does it remain at constant distance?). However, even without range, lines of sight (LOS) can be reconstructed to provide potential flight paths as a function of range, and its temporal evolution. The lines of sight, or lines of bearing, indicate where the ATFLIR targeting pod is looking at any moment in the video (i.e., in which direction the object is relative to the F/A-18F).

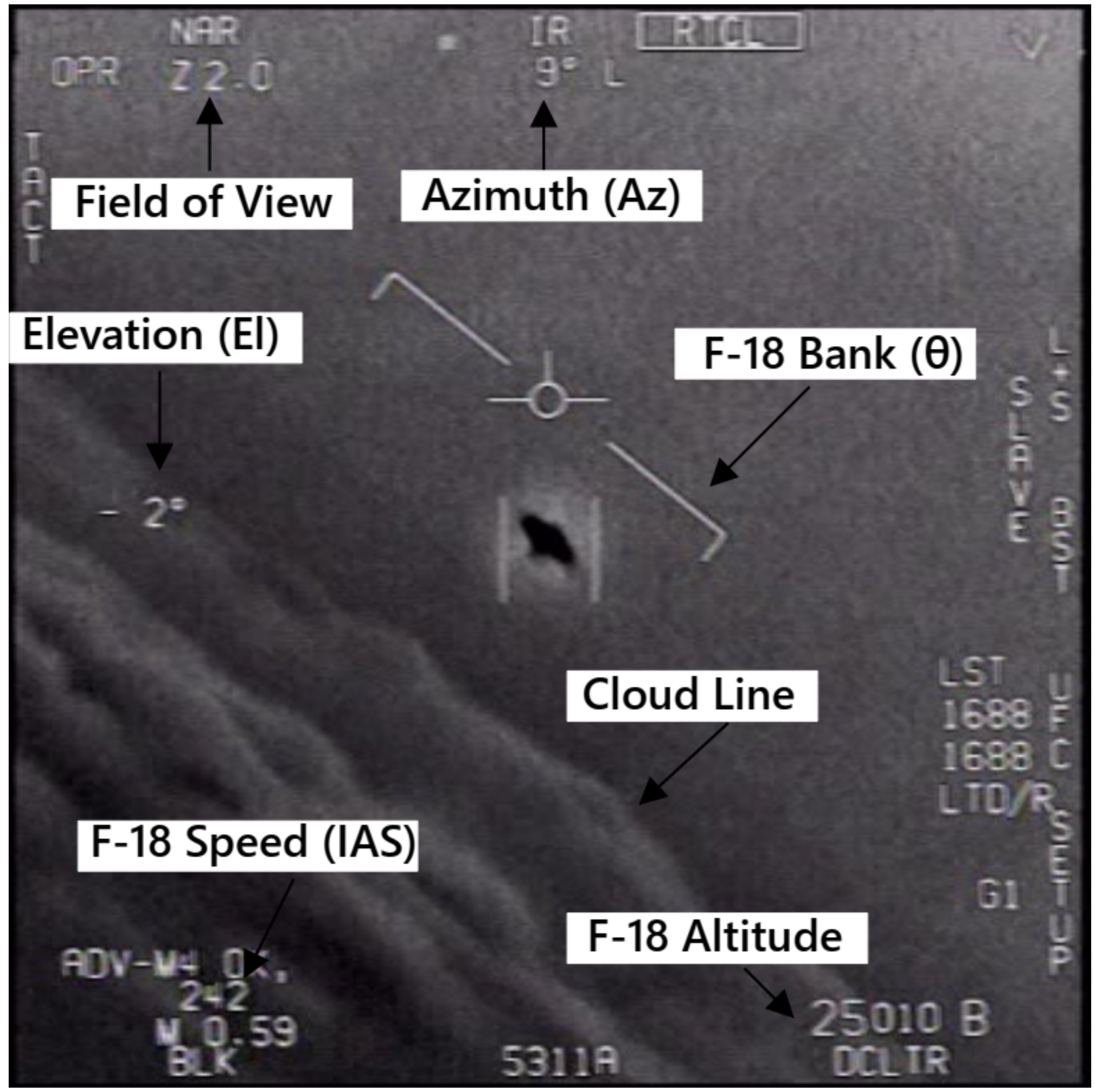

**Figure 2. ATFLIR display and parameters of interest for potential flight path reconstructions.**

### A. F/A-18F flight path

A prerequisite to reconstruct the LOS is to estimate the F/A-18F flight path. This can be done using two parameters provided on the ATFLIR display: Aircraft speed and aircraft rate of turn (ROT).

- *Aircraft Speed:* The ATFLIR display provides the indicated air speed (IAS) of the F/A-18F in knots (kts) at any moment; it is displayed in the bottom left of the screen (Figure 2). IAS equals true air speed (TAS) at sea level but is less than TAS at altitude because it accounts for the decrease in atmospheric density, an important parameter to consider for flight performance. TAS is the actual speed of the aircraft relative to the air mass in which it is flying. It is different from ground speed - the speed of the aircraft relative to the surface - which depends on the wind (i.e., speed of the air mass relative to the surface). TAS is required to reconstruct the F/A-18F flight path. It can be derived from IAS using a simple TAS calculator.[8] IAS varies between 238 and 242 kts throughout the video. At 25,000 feet (ft), and with an air temperature of ~ -25°C (estimated from meteorological data around that altitude on January 24th, 2015), this corresponds to a TAS that varies between 363 and 369 kts. For each selected frame (see below), the corresponding TAS value from that interval is used in the calculations.

- *Aircraft Rate of Turn (ROT):* The rate of turn corresponds to the number of degrees of heading change per second. The exact ROT is unknown but can be estimated from the plane's bank angle relative to the horizon (Figure 2). The bank angle has been extracted every 10 frames of the video. Because the frame rate, or frames per second (fps), of the video is 30 fps, the video consists of 1034 frames. This gives 103 values of banking angle, from frame 1 to 1031. The rate of turn in °/s is derived from the following formula commonly used in aviation [2]:

$$ROT = 1091 \times tan(\theta) / TAS \quad \text{with } \theta\text{: banking angle; } TAS\text{: true air speed} \qquad (1)$$

The F/A-18F air track (trajectory relative to its surrounding air mass) can be estimated every 10 frames using the speed and ROT. Ground track (trajectory relative to the Earth's surface) requires accounting for wind at the location of the F/A-18F, which is unknown. However, the audio from the Gimbal video gives an indication, as the WSO states that the wind is "120 knots to the west" (i.e., the wind is westerly/blowing from the west, with a speed of 120 kts). Knowing the exact date and time for the Gimbal incident could confirm wind speed and direction, but the relevant information has been contradictory. While Department of Defense statements[9] point to January 21, 2015, the metadata of the video officially released by DoD indicates that it was encoded on January 25, 02:29 UTC. This corresponds to January 24, 9:29 pm local time (Eastern Standard Time), and would be consistent with a video digitized on the USS Roosevelt aircraft carrier following an evening training flight (LT Graves has stated that Gimbal was filmed after sunset). We have also obtained an unclassified Department of Defense email that dates the event to January 25. Using hourly wind data from the ERA5 reanalysis [3], we find that the evening of January 24 is a better match for the 120-kt westerly wind than either evenings of January 20 or 21, 2015. A strong jet stream was present off the coast of Jacksonville, FL, that corresponds to a westerly wind of around 110-120 kts around the 25,000 ft altitude of the F/A-18F (Figure A1).

Accounting for the wind when reconstructing the F/A-18F flight path is critical, because the speed/ROT seen in the video correspond to significantly different flight paths depending on the wind. The F/A-18F flight path is calculated using the following equations of motion to determine its coordinates $x$ and $y$ at any time $t$, every 10 frames:

$$x(t) = x(t-1) + (TAS.\Delta t.sin(\beta)) - (W_s.\Delta t.sin(W_d)) \qquad (2)$$

$$y(t) = y(t-1) + (TAS.\Delta t.cos(\beta)) - (W_s.\Delta t.cos(W_d)) \qquad (3)$$

With : $TAS$ true air speed in Nm/s

$\Delta t$ time interval between 10 frames (~0.3 s)

$\beta$ heading angle in ° (starting at 0°)

[8] https://e6bx.com/tas/

[9] https://www.theblackvault.com/documentarchive/u-s-navy-releases-dates-of-three-officially-acknowledged-encounters-with-phenomena/

$W_s$ wind speed in Nm/s

$W_d$ wind direction (0° for headwind, 180° for tailwind)

The ATFLIR display indicates that the F/A-18F flew at an almost constant altitude of 25,000 ft, so the flight path can be reconstructed in a two-dimensional plane.

Figure 3 shows the reconstructed flight path using equations (1) and (2) above, with no wind (dashed line) and with a 120-kt headwind from 35° to the left of initial heading (solid line). As the F/A-18F banks to the left, as indicated by the horizon indicator in the video, it gradually turns to the left. With wind included, its direction of flight is deflected and it finishes its course around coordinates (-1. ; 1.9), in Nm, at the end of the video. The effect of wind is significant, and must be accounted for when reconstructing flight paths. However, because the elevation angle of the ATFLIR pod is only -2°, objects at relatively close range remain at an altitude that is close to 25,000 ft (i.e., the wind speed and direction are roughly the same for a nearby object and the F/A-18F). Thus, the effect of wind is similar for both, so that its influence can be neglected and the reconstructed paths are representative of paths in the same moving air mass.

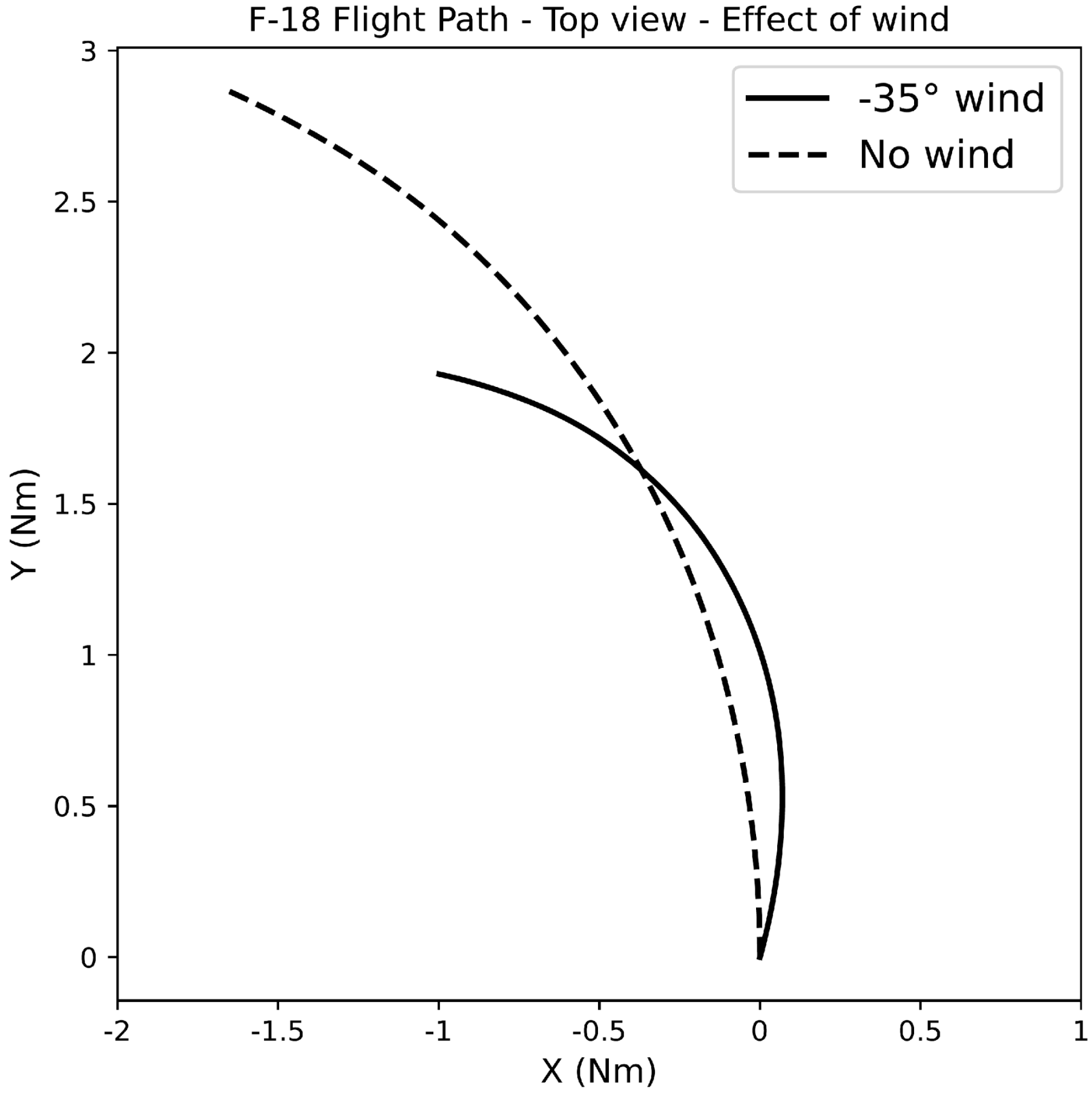

**Figure 3. Reconstruction of the F/A-18F flight path at 25,000 ft, based on true air speed (~365 kts), banking, and estimated wind, over the 34 seconds of the video. The F/A-18F starts from the (0,0) origin. The solid curve represents the flight path for a 120-kt wind blowing from -35°, relative to initial heading. The dashed curve is the flight path in the local air mass.**

## B. Lines of sight (LOS)

Once the F/A-18F flight path has been reconstructed, the lines of sight can be estimated using the azimuth (Az) and elevation (El) angles indicated on the ATFLIR display (Figure 2). Because the Az values are rounded and updated every other frame, they need to be smoothed to get continuous values every 10 frames. This is done using a moving average of 50 frames, based on the raw Az values extracted from the video. In addition to smoothing the Az angle, we also take advantage of the background cloud motion to refine the LOS. The Az angle is rounded to integer values and it is relative to plane boresight, which is not aligned with direction of travel in the presence of crosswind. This creates uncertainties in the exact Az relative to the F/A-18F flight path. In the video, about 9 fields of view (FOV) are scanned by the camera. The field of view (FOV) of the ATFLIR pod in NAR2 mode is 0.35° (0.35°x 0.35° along the horizontal and vertical), magnified twice with digital zoom from the NAR (narrow) 0.7° optical FOV.[10] This means about 9×0.35=3.15° of the sky are effectively scanned by the camera, as most of the change in Az (from 54°L to 7°R) occurs to compensate for the plane's turn. This is useful information that greatly helps to get smoother LOS and more consistent results (discussed further at the end of Section IV.A). Az is refined by adding a term in the calculation of the LOS, that gradually adjusts them to match the change in number of scanned FOV as observed in the video.

El is also rounded and stays constant at -2° during the video. The exact rounding method is unknown, but -2° likely corresponds to a [-1.5°/-2.5°] range for El. Assuming a flat cloud cover (an important assumption), the El angle does not change significantly because the cloud line remains in the field of view. If El was changing by a few tenths of degrees, the cloud line would move significantly and disappear from the FOV. But it does not, and we only see the cloud line lowering slightly in the FOV, by about ⅛ of the FOV, or about 0.05°. To account for this and constrain the El angle, we start from an El angle of -2°, that gradually increases to -1.95°, linearly across the 104 reconstruction points. The constraint El based on the cloud line significantly constrains the LOS and potential flight paths, since an object getting closer to the F/A-18F must climb in altitude to remain in the FOV (the converse is true for an object moving further away).

The lines of sight that we retrieve utilizing the method described above are shown in Figure 4, within 10 Nm from the F/A-18F. They show directions where the object may potentially be relative to the F/A-18F, every 10 frames (0.3 second) of the video. The F/A-18F trajectory here does not account for wind, so this assumes that the wind is similar for both the F/A-18F and the Gimbal object. Within 10 Nm, the object remains above 23,000 ft, where the wind is usually not very different from 25,000 ft, so this is a reasonable assumption for a close-object flight path.

The object is on the left of the F/A-18F (initial Az of 54°L), then as the F/A-18F turns towards it (leftward bank), the Az decreases and around frame 910 the object is in front of the F/A-18F's nose (Az=0°), before passing to its right (Az=7°R at the end of the video). Most motion observed in the video is from the parallax effect, (i.e., the apparent motion of the object relative to the background clouds) as a result of the F/A-18F's displacement. From the top-view perspective of Figure 4, we observe that the distance between the lines of sight decreases with time, which is consistent with reduced angular speed of the clouds over the course of the video. The lines of sight become almost indistinguishable from each other around frame 900, then they reverse direction in a curve along the vertical (which may be described as a "breaking wave"), as discussed later and better seen in Figure 8. Before further investigation, we can already note that this geometry of the lines of sight is consistent with aircrew accounts of an object reversing direction at that range, when seen from above, as on a SA radar display, and in Figure 4.

---

[10] https://web.archive.org/web/20091211103559/http://www.raytheon.com:80/businesses/rtnwcm/groups/sas/documents/content/rtn_sas_ds_atflir.pdf

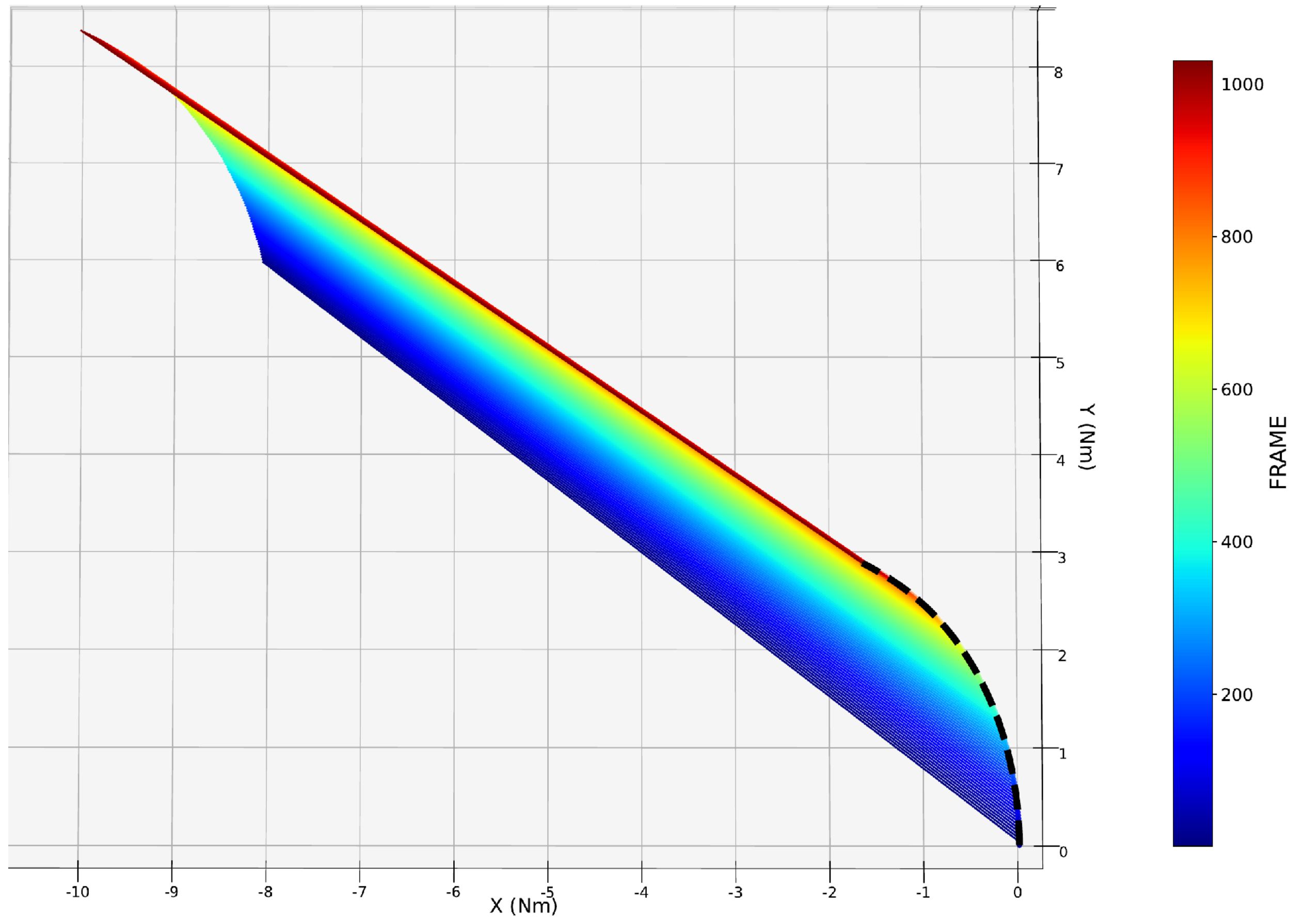

**Figure 4. F/A-18F path (dashed black curve) and lines of sight (colors, top-view), in function of time (video frame). The lines of sight have been adjusted to match angular cloud motion observed in the video. The wind effect is not included in the F/A-18F flight path to highlight potential trajectories in a "same-wind" frame of reference.**

## IV. Results

### A. Potential flight paths within 10 nautical miles and the "vertical U-turn"

Once the lines of sight are reconstructed, we can estimate potential flight paths for the object as a function of distance. In this section, we want to examine flight paths as they would have been seen on the SA display (i.e., relative to the ground), so we include the effect of wind. Mathematically, there is an infinite number of potential flight paths along the lines of sight. However, direct data from the video, as well as second-hand data provided by naval aviators (noted below), can be used to constrain the range of potential flight paths.

- The Gimbal object was within 10 nautical miles [1], with a finer estimate of 6-8 Nm provided by the WSO via LT Graves.

- The object was traveling against a westerly 120-kt wind, as stated in the video. Note: This is consistent with the WSO stating that the object was first detected coming from the east and heading towards the USS Roosevelt aircraft carrier (Figure1).

- The object was traveling in a straight line, stopped momentarily, then reversed direction without radius of turn, before the rotation observed in the FLIR video[11] [1].

- The apparent size of the object on the FLIR grows by ~15%. This was verified through an image analysis of the object's size (Figure A2), at the beginning versus end of the video (accounting for the fact that apparent size is larger in "black-hot" mode, relative to "white-hot" mode). Given that apparent size is a function of the invert of distance, this corresponds to a decrease in distance of ~ 13% (1/1.15~0.87).

With this information, we build a potential flight path that starts at 8 Nm, proceeds in a straight line, and gets closer by ~13% to match the change in apparent size, while facing a 120-kt wind. At 8 Nm, this corresponds to an approximate size for the object's IR signature of 15-20 ft, along its longest axis, considering the object spans about $1/15^{th}$-$1/20^{th}$ of the horizontal 0.35° FOV (see formula in Appendix). It is important to remember that what is seen in the Gimbal video is an infrared signature, and how it reflects the true shape and size of the object is uncertain. But as stated by LT Ryan Graves and suggested in Figure 1, the F/A-18F came closer to the object during the intercept, so it is logical to examine flight paths for which the distance to the object decreases (and apparent size of the object increases). To estimate wind direction, we can use the change in apparent size. Straight flight paths corresponding to different wind directions, as defined relative to the F/A-18F initial heading, with different offsets for the direction of Gimbal relative to wind, are tested in the lines of sight. The corresponding change in distance to the F/A-18F, from start to finish, is measured. The results are shown in Figure 5. The x-axis shows possible wind directions, from -70° to 70° relative to the F/A-18F initial heading. This range is chosen based on the geometry of the lines of sight (Figure 4), and the constraint that the object must cross the lines of sight while going roughly against the wind (audio). In Figure 5, the colored, dashed/solid curves represent different directions for the Gimbal object, defined as an offset relative to the specified wind direction. This accounts for the fact that the object may not fly exactly against the wind. The range for this offset goes from -40° to 40°, sufficient to remain consistent with an object observed flying against the wind on the overhead radar display. For all these combinations of wind direction and offset, the change in distance for an object proceeding in a straight line through the lines of sight is measured on the y-axis, in percent (%) of the initial distance (8 Nm). The yellow shaded area marks the estimated change in distance in the video, from the measured change in apparent size of the object. Our image analysis suggests a change of ~ -13% in distance. We show the -10/-20% interval here to account for uncertainty in this estimate.

A "sweet spot" is found for the simulated versus observed (estimated) change in distance, for wind directions between -50° and -15° (relative to the F/A-18F's initial heading). In that interval of wind direction, the different curves converge in the interval of observed estimated change in distance (yellow area). Based on this analysis, we select a wind direction of -35° (close to the center of the wind direction interval), and an offset of 15° (that sits in the upper section of the shaded area, for that wind direction), to highlight a potential flight path for the object, in the range provided by the aviators, in the next figure. On a side note, in this range of wind direction, the F/A-18F must experience a crosswind, to its right, towards the end of the video. After close investigation of the video, we note that the real horizon (cloud line) and the artificial horizon (plane bank in Figure 2) are misaligned for much of the video. This effect is expected when looking at the horizon from a plane in a bank. It seems that the two horizons realign between Az=20°L and 10°L (between frames 500 and 800), before the camera points at Az=0°. Although more speculative given the potential influence of other factors like error in the ATFLIR derotation mechanism, we expect both horizons to be perfectly aligned when the camera points in the direction of travel. This observation seems consistent with the plane "crabbing" in a right crosswind in the last section of the video, which is expected in our selected range of possible wind direction.

---

[11] https://thedebrief.org/devices-of-unknown-origin-part-ii-interlopers-over-the-atlantic-ryan-graves/

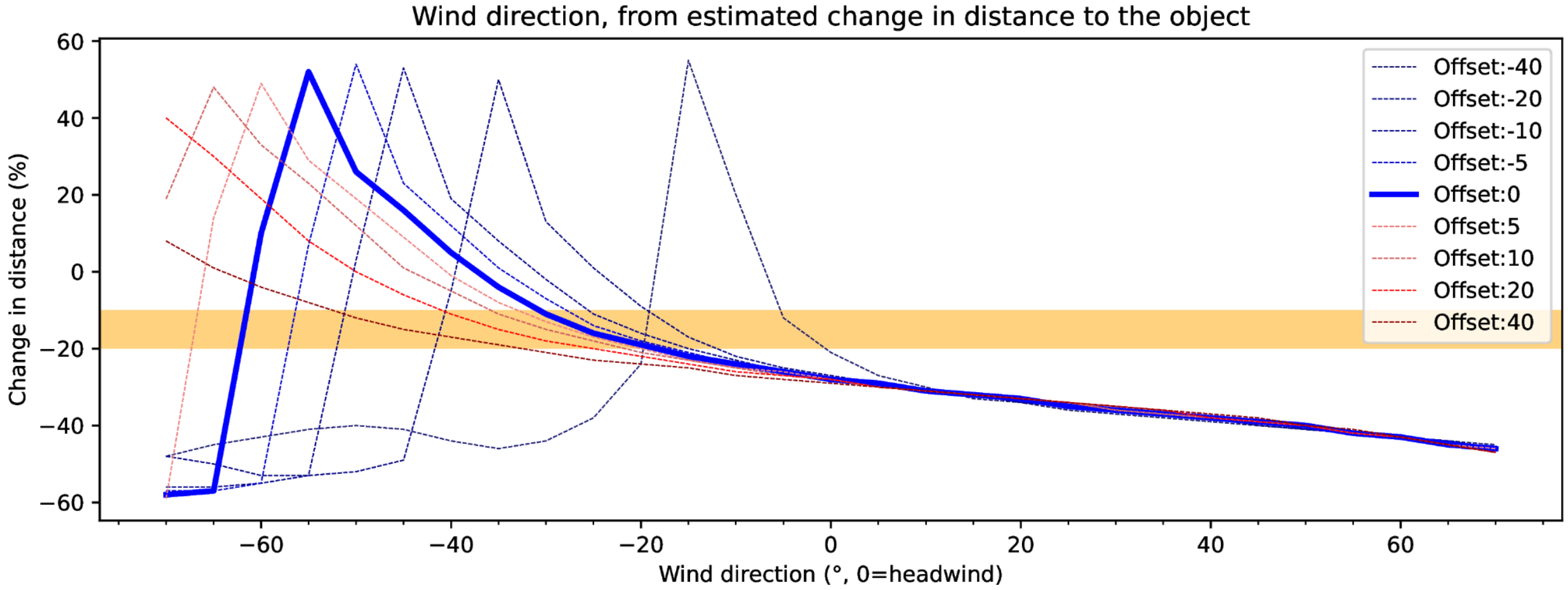

**Figure 5. Change in distance of the object to the F/A-18F, from beginning to end of the video, for different wind directions (x-axis, relative to initial heading) and offset in the Gimbal versus wind directions (curves, relative to wind direction). This is for straight paths through the lines of sight. The yellow shaded area marks the expected change in distance when estimated from the observed change in apparent size of the object in the video. Negative (positive) wind direction is to the left (right) of initial heading.**

Figure 6 shows a potential flight path that obeys all the conditions listed above: The object travels in a straight line, starting at 8 Nm. It generally travels against a 120-kt wind that is oriented 35° to the left of the F/A-18F initial heading, and the F/A-18 is ~13% closer to the object at the end, versus the beginning, of the flight path (6.8 Nm end distance). In this configuration, and as shown in Figure 3, the F/A-18F flight path is condensed and deflected to the left, compared to the flight path that does not account for wind. Accounting for the effect of wind, Figure 6a shows the object's flight path relative to ground, as it would be seen on the SA display. In this frame of reference, the object needs to reverse direction around frame 800 to follow the folding of the lines of sight (Figure 6b). The object then proceeds in the opposite direction until the end of the video.

Since this trajectory is relative to the ground, it can be directly compared to the anomalous flight path observed by the aviators on the SA (radar) display. Figure 7 shows the corresponding ground speed of the object in that particular scenario (green curve). The "true" air speed of the object, in its local air mass, is shown in blue. The object starts at an airspeed of about 400 kts, which decreases almost linearly to under 50 kts around frame 900, about when the long, continuous rotation occurs in the video (marked by the yellow shaded area). After this rapid slowdown and near-stop, the object slowly regains speed. Because the object is going against the wind, ground speed is lower (green curve). It starts at about 300 kts and reaches a minimum, close to 0, around frame 720. At that moment, the object still has intrinsic speed against the wind but, because it passes under 120 kts relative to the ground, the object starts traveling in the opposite direction for a ground-based observer. This moment may correspond to the instant when the aircrew and Navy personnel who watched the tapes observed a brief stop on the SA, and a 180° reversal of direction with no radius of turn. Interestingly, this instant coincides with the first step in the counterclockwise rotation of the object observed in the video (first vertical orange dashed line in Figure 7, observed around frame 720). This may suggest that the object was somewhat taken by the wind after decelerating under wind speed (120 kts), and that it started tilting while being picked up by the wind flow. As seen in the air speed graph, the object still has motion against the wind until about frame 900, when a longer rotation of the object occurs on the FLIR. This suggests that the rotation is associated with a change of direction, as now the object is traveling with the wind, slowly speeding up in the wind flow (which explains a ground speed greater than 120 kts by the end of the video).

The comparison of air versus ground speed in this scenario is particularly intriguing when placed in context of LT Graves' recollection of the events, as seen on the FLIR and SA displays.[12] Graves states that the object stopped on the SA *before* tilting up as seen in the video, and proceeding in the opposite direction. This is consistent with the scenario we describe here, as the effect of facing the wind would have made the object exhibit a stop/reverse on the SA display a few seconds before it reverses direction in its own air mass. Note that initial air speed is quite uncertain as it depends on the angle through which the object traverses the lines of sight (i.e., offset direction from wind), as well as initial distance. A reasonable range for initial air speed is 200-450 kts. The emphasis here is on the decrease in air speed below wind speed and the reverse of direction; both results are robust within 10 Nm in the range of wind direction/offset shown in Figure 5.

As further discussed in Section IV.B, the question of what is causing the rotation of the Gimbal object is controversial. One theory proposes that the rotation is simply an optical artifact of the gimbal-mounted ATFLIR camera that needs to roll around Az=0° to maintain track of the object. However, in the scenario that considers the context and aviators' accounts, there is a remarkable coincidence between the moment when the object rotates in the video, and the curve of the lines of sight in the local air mass (breaking wave, Figure 4). This is better seen in Figure 8, which shows the same trajectory as in Figure 6, but in the local air frame of reference rather than ground frame of reference (assuming wind is the same for both the object and the F/A-18F). This "real" trajectory of the object describes an object traveling in a straight line against the wind (by design), climbing by a few hundred feet (see Figure A3 for the ~250 ft altitude increase over the course of the video) and reversing direction along the vertical, in what can be described as a "vertical U-turn," or a "J-hook" trajectory. The vertical U-turn coincides with minimal air speed in Figure 7 and, as noted above, coincides with the long, continuous rotation of the object in the FLIR video.

Altogether, the trajectory we highlight here is consistent with aircrew accounts of an object traveling against a westerly wind before reversing direction (Figure 1). In that scenario, the object slows down, comes to a stop relative to the ground (SA display), is taken by the wind after passing under 120 kts, starts tilting, and executes a vertical U-turn to reverse direction while tilting further. It finally gains speed in the wind flow and moves eastward. This all occurs while the F/A-18F is closing on the object.

It is important to remember that we only present one potential trajectory here, which depends on the assumptions listed at the beginning of this section. However, any straight-line flight path within 10 Nm that goes against the wind through the lines of sight requires a reverse of direction along the vertical to follow the LOS. Changes in initial distance and heading affect the amount of initial speed the object has, the amplitude of its altitude climb, and exact timing for minimum ground and air speed. For example, loosening the "straight path" constraint allows for a horizontal component to the final U-turn (i.e., a slight horizontal radius of turn). But qualitatively, at that range, the solutions do not change drastically and they conform with aviators' accounts of an object stopping and subsequently reversing direction with no or low radius of turn. The scenario that we detail in this section simply represents our "best-guess" estimate, constrained by the change in apparent size of the object. Still, the results are not extremely sensitive to exact parameters that are chosen, as long as they obey the general configuration of the event (object traveling in a straight line against a 120-kt wind, within 10 Nm).

One of the most uncertain parameters of this reconstruction is the change in altitude, largely because we do not know small variations in the elevation angle of the pod, and that we impose a linear increase in this angle to match how the clouds drop slightly in the FOV. This likely results in smoothing actual potential changes in altitude that may happen less progressively. In fact, close investigation of the distance between the cloud line and the object reveals an intriguing result. While the lowering of the cloud line in the FOV, relative to the object, looks smooth and gradual, there is an apparent jump of the Gimbal object around frame 880. As illustrated in Figure 9, if we select frames 821, 881 and 941 (2 and 4 seconds apart), and stitch the frames together to align the clouds in a horizontal plane, we see that the object moves slightly higher above the clouds during this time interval (see orange line, parallel to cloud line in green). This is also seen by the velocity vector and artificial horizon climbing progressively relative to the clouds (see red line, Figure 9). Because this coincides with the long, continuous rotation of the object, it may be evidence that the object is executing the vertical U-turn at that specific moment.

[12] https://thedebrief.org/devices-of-unknown-origin-part-ii-interlopers-over-the-atlantic-ryan-graves/

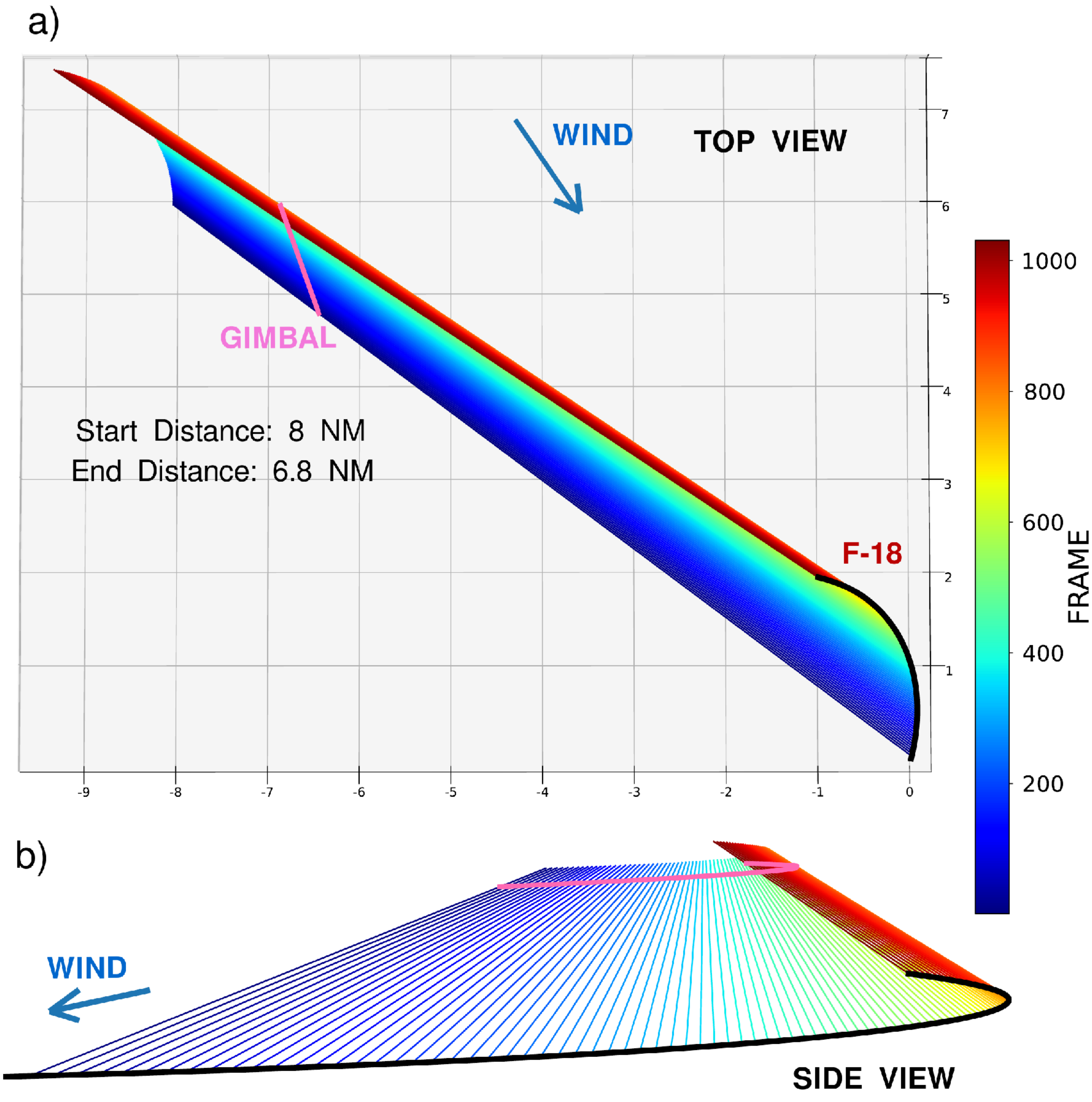

**Figure 6. Potential flight path (pink curve) for an object traveling in a straight horizontal trajectory, starting at 8 Nm, ending at 6.8 Nm, from the F/A-18F (black curve). In this scenario, the object gets 13% closer, consistent with the ~15% increase in apparent size in the video. a) Top view. b) Side view. The "vertical U-turn" towards the end is consistent with the object stopping momentarily on the SA (around frame 720 in this scenario). The z-axis is scaled with a factor 2 to highlight the vertical motion. The blue vector indicates the 120-kt wind direction for this scenario (35° to the left of initial F/A-18F heading). The object has an offset of 15° compared to wind direction (i.e., its direction is oriented 20° to the left of initial F/A-18F heading).**

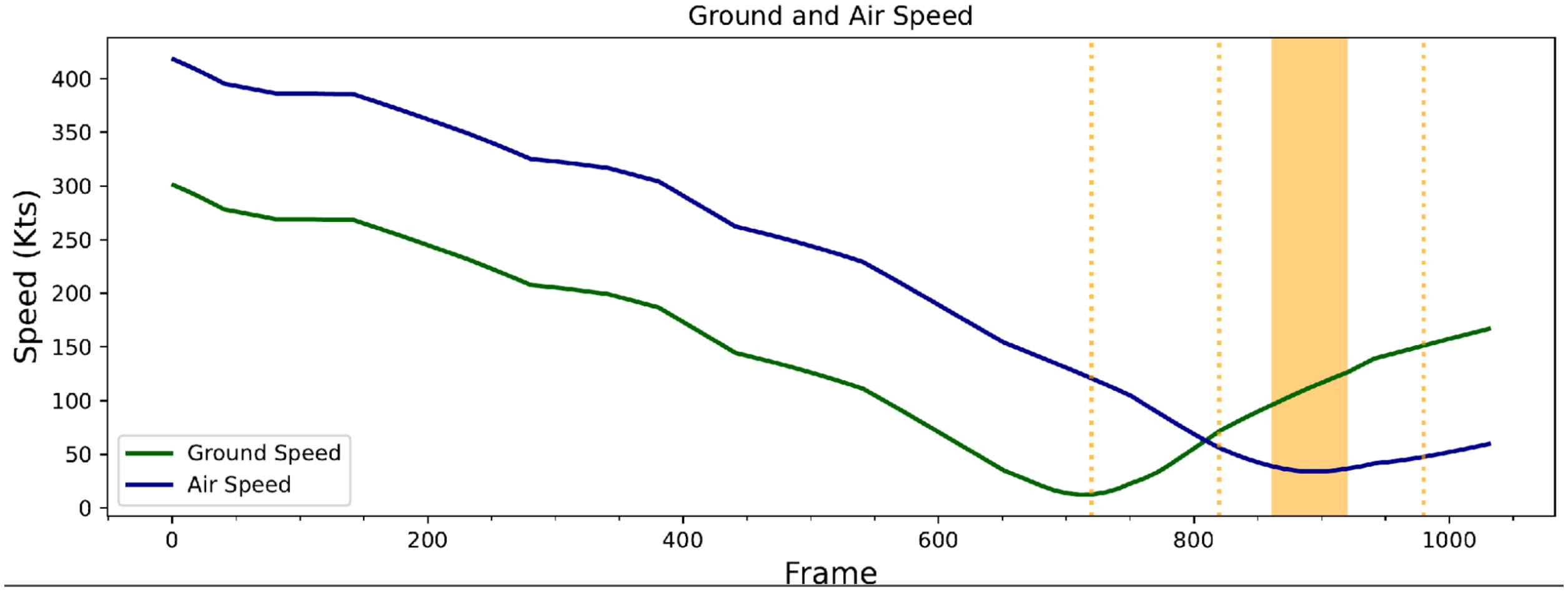

**Figure 7. Ground speed (green) and air speed (blue) of the Gimbal object throughout the video, in the scenario shown in Figure 6 (wind direction=-35°, offset=15°, initial distance=8 Nm). Unit is knots. The yellow shaded area marks the long, continuous rotation in the video. The vertical dashed lines mark the brief "step" rotations around frames 720, 820, and 980. Note that initial speed varies quickly with small changes in flight path parameters (wind direction, offset, initial distance). For example, a very similar flight path but with an initial ground speed of about 180 kts is found with slightly different values of these parameters (wind direction=-25°, offset=20°, initial distance=9 Nm).**

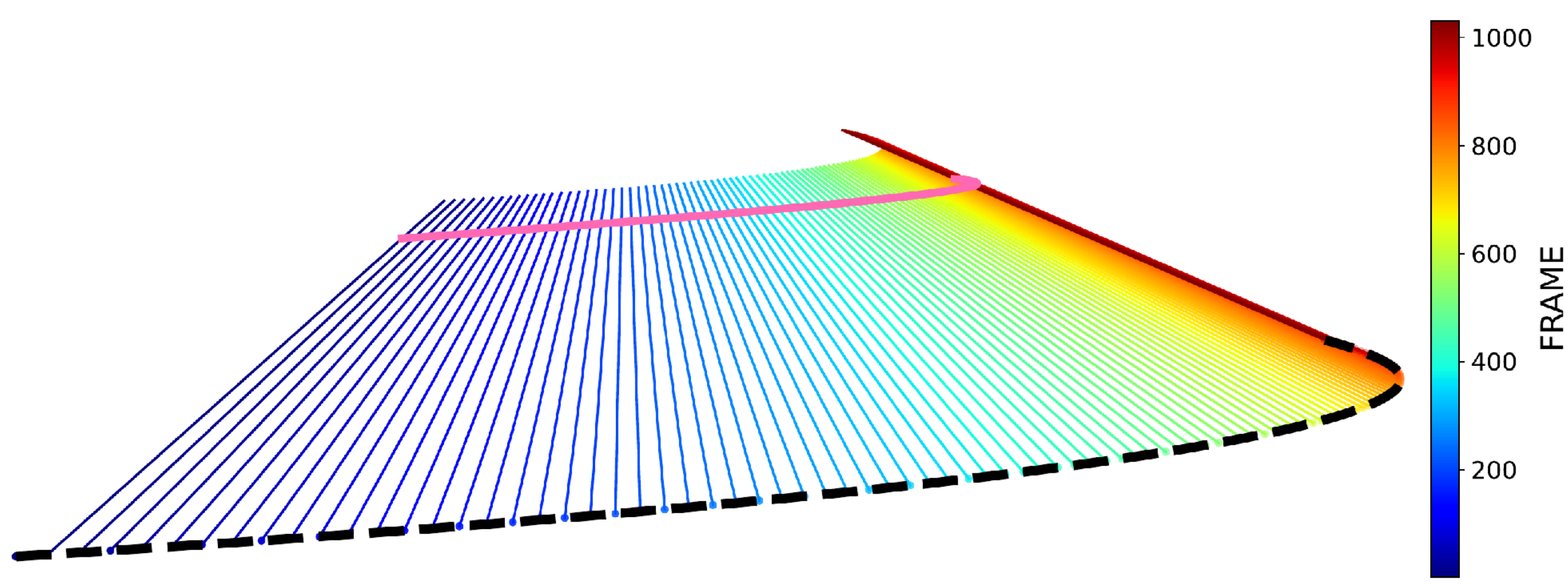

**Figure 8. Potential flight path (pink curve) of Figure 6 in the referential of the object (no wind). The z-axis is scaled with a factor 2 to highlight the vertical motion.**

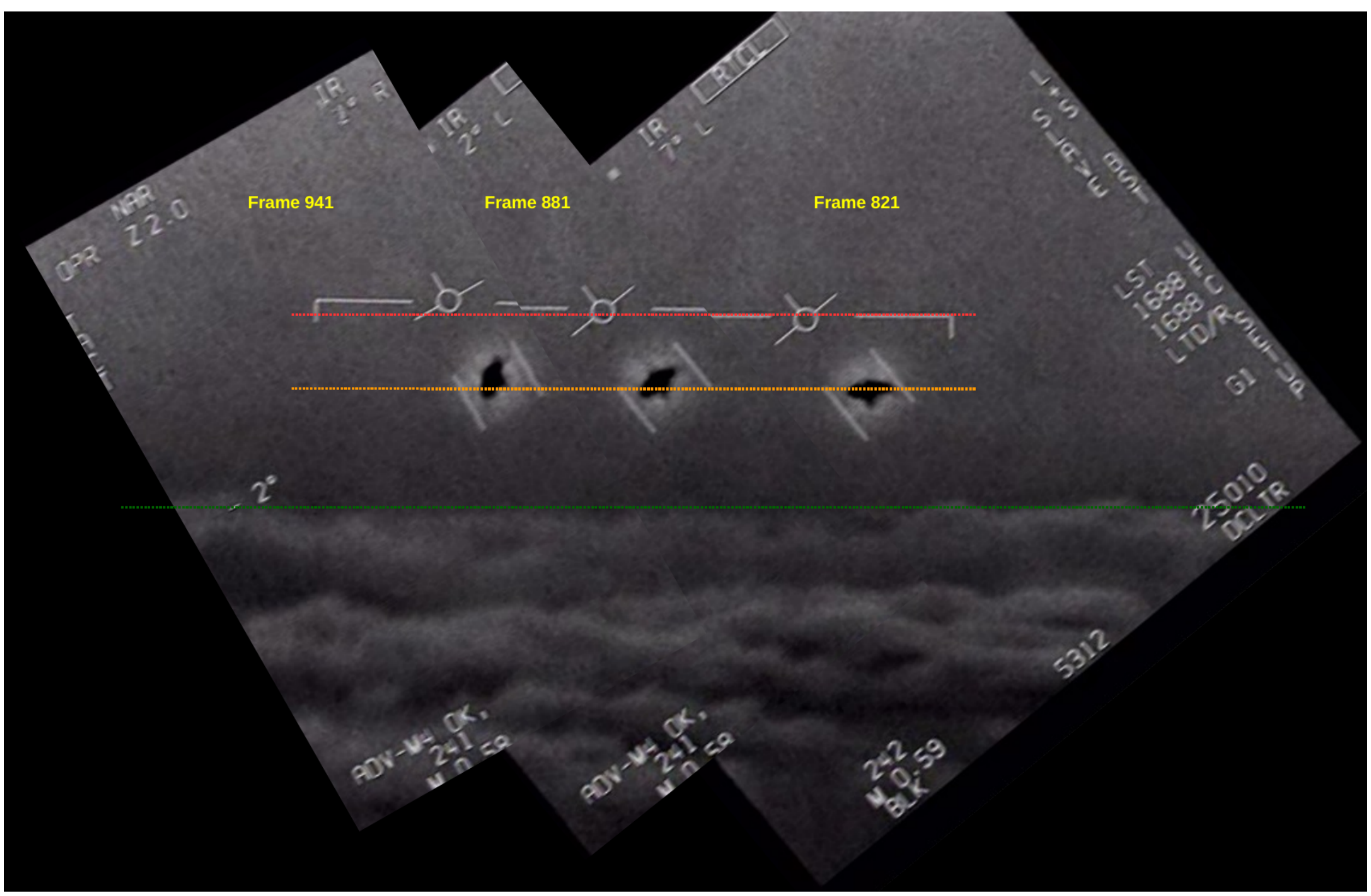

**Figure 9. Possible altitude increase (as seen in the ATFLIR FOV).**

Due to contradicting information in the available documentation on the ATFLIR pod, there has been debate among investigators on whether the ATFLIR NAR2 field of view is 0.35° or 0.7°. The current consensus is that the NAR2 FOV is indeed 0.35°, but to illustrate that it does not strongly affect our results concerning the potential close flight paths (within 10 Nm), Figure A4 shows the same flight path as Figure 6 but with a FOV of 0.7°. The choice of FOV affects the adjustment of the LOS using background cloud motion (Section III.B). As can be seen in Figure A4, using a FOV of 0.7°, the flight path is relatively close to the flight path of Figure 6, with a similar vertical U-turn towards the end of the video. The speed of the object is reduced, starting at 250 kts, and it climbs a bit more, by 350 ft. In Figure A5, we also show the results when no adjustment of the LOS using cloud motion is applied. Results are noisier, although qualitatively similar in the overall trajectory (direction reversal, as described by witnesses, via a vertical U-turn). We believe that the movement of the clouds is an important piece of data in the video, so we favor adjusting the LOS using that information, as done in our previous analyses. For completeness, and to emphasize that the main conclusion concerning flight characteristics within 10 Nm (stop/reverse along the vertical) are robust across a wide range of scenarios, we also show another potential flight path in Figure A6. This one is for a wind direction oriented 35° to the right of the F/A-18F (offset for the object direction equals 10°), which corresponds to a 90° crosswind affecting the F/A-18F around the middle of the video. Although this scenario also verifies the "vertical U-turn" trajectory, the F/A-18F gets closer to the object faster, which should correspond to an increase in apparent size of 50% (not observed in the video). The object also needs to climb by 600 ft, which seems less realistic than the flight path of Figure 6. But this is another trajectory in the range of potential flight paths within 10 Nm.

### B. Discussion of the rotating "glare" and "distant plane" hypothesis

UAP investigator Mick West has developed a three-dimensional simulator, referred to as Sitrec ("situation recreation") which aims to reproduce the Gimbal encounter.[13] To test varying scenarios, many key parameters are adjustable: F/A-18F speed, initial distance of the object, type of trajectory (straight line, constant speed, constant altitude, etc.), wind speed/direction for the F/A-18F, object, clouds, etc. In addition to representing the F/A-18F flight path, the LOS, and potential trajectories for the object in three dimensions, the simulator emulates the ATFLIR display and cloud motion, which helps ensure the reconstructed LOS are consistent with cloud motion observed in the video.

For close trajectories within 10 Nm, such as the one we present in Section IV.A, Sitrec and our own recreation are remarkably consistent. They are also consistent with similar efforts made by two independent researchers who have investigated this case (see Acknowledgements). This is expected since similar methods for constraining the LOS have been used by us, Mr. West, and these researchers. Nevertheless, it gives confidence in the results from the present study since the general flight paths at the various ranges are corroborated by four independent models. Within the 10 Nm distance consistent with aircrew accounts, and accounting for the elevation angle constraint, straight-line trajectories for the object describe a "vertical U-turn" and a reverse of direction towards the end of the video. Figure 10 shows a straight-line flight path in Sitrec, with an initial distance of 9 Nm, and the heading of the object's trajectory adjusted so that the change in apparent size of the object in the simulation is roughly consistent with change in size in the video. As in our best-guess estimate of Section IV.A, the wind is set to 120 kts, blowing from 35° to the left of the F/A-18F's initial heading. The object roughly faces the wind, in agreement with the audio, with an offset from wind direction close to the offset of 10° we selected in Section IV.A. The left graph in Figure 10 shows the object's ground speed in green, and its air speed in blue. The right graph indicates the change in altitude, from the beginning to the end of the simulation. As in our reconstruction, the object slows down under wind speed and takes a sharp turn with little or no horizontal component (i.e., a "vertical U-turn"). Again, this flight path would be seen on the SA display as a stop then reversal of direction, since the ground speed reaches 0 around frame 780.

UAP investigator Mick West does not believe this scenario is plausible and offers an alternative hypothesis. As seen in Figure 10 (yellow line), in Sitrec the LOS flatten at about 30 Nm from the F/A-18F, such that an approximately straight-leveled trajectory can be drawn at that distance. We also find this solution, although we do not find a particularly straight nor steady flight path (Figure A7). These solutions are a consequence of the lines of sight diverging in the distance, while they converge within 20 Nm, such that leveled paths that may resemble a plane trajectory can be retrieved around 30 Nm (although we find that to remain within the LOS and maintain constant altitude, air speed must increase from 300 to 450 kts over the course of the video). Mr. West proposes that this is the most likely trajectory for the Gimbal encounter and that it is consistent with another plane observed more or less "tailpipe" (i.e., tail-on from behind, looking at the engines). According to Mr. West, the IR signature observed in the video is an infrared "glare" induced by engine exhaust. In this scenario, the rotation observed in the video cannot be intrinsic to the object, so this scenario hypothesizes that it is caused by the ATFLIR pod rotation, which needs to roll around Az=0° to maintain lock and avoid the singularity/gimbal lock. Proponents of this scenario argue that the name of the officially released video, "Gimbal," is a reference to the anomalous rotation of the object. This theory has been widely shared in the public as a potential explanation for the Gimbal UAP, and it has been reported that some anonymous military officials believe that the object's anomalous motion is caused by the ATFLIR pod optics.[14]

We do not intend to thoroughly discuss the plausibility of this scenario in the present paper. However, we mention it in order to provide a complete overview of the status of investigations on this case. Indeed, it is a hypothesis that has been defended by Mr. West and others since 2017, with detailed analyses. We will simply mention what we believe are difficult-to-reconcile flaws in the "distant plane" scenario:

1. It supposes that the range of the object provided by witnesses is wrong, and that the F/A-18F aircrew made an error by locking onto the wrong target (a "random" plane in the distance). Beyond the fact that this aligns very poorly with what is described in the cockpit audio and the official report of the incident (Figure 1, "That [the ATFLIR target] *is* the L+S [the radar target]", "stable trackfile", "closest point of intercept", etc.), the probability of erroneously locking a distant plane, 30 Nm away, in a tail-on alignment, in the FOV of 0.35°, while investigating an unrelated anomalous radar return, seems rather low; especially if aircrew receive an indication that the ATFLIR is slaved to a designated radar contact ("That [the ATFLIR target] *is* the L&S [the radar target]").

2. It supposes that the Navy was unable to identify a plane flying in a tightly-controlled training range during a pre-combat deployment exercise involving a carrier strike group (i.e., onboard and outboard radar data). (Personal communication with LT Ryan Graves)

3. The "fleet" of objects must be independent of the Gimbal object, so this scenario requires anomalous radar returns coinciding with an erroneously locked target.

4. More difficult to explain, the LOS from the supposed distant plane would mimic the close trajectory within 10 Nm (Figure 6), as described by the aircrew (i.e., abrupt stop and reversal of direction after flying against the wind). This suggests a link between what was seen on SA and the distant plane locked by error, which can only be reconciled with radar errors and/or spoofing (discussed in Section IV.C).

5. Assuming that the potential distant plane is another F/A-18F involved in the exercise, or even an airliner, the shape and size of the IR signature seems disconnected from the geometry of the heat source (Figure A8). Examples of ATFLIR footage are limited so it is difficult to estimate how anomalous an IR signature this would represent (see Figure A9 and A10 for some publicly-available examples of such footage). But to our knowledge, no fighter pilot has publicly supported the idea that the infrared image in Gimbal is consistent with the exhaust of an aircraft some 30 Nm in the distance.

6. This theory supposes that the roll motion of the ATFLIR pod occurs in steps (as seen in the Gimbal video, the object does not rotate smoothly/continuously). This requires that internal mirrors inside the pod compensate for the absence of pod roll over a few degrees (2-3°) before the pod rolls abruptly to re-center the object in the optical system, every time the internal mirrors cannot compensate for deviation of the target from the center of the optics (see the ATFLIR pod simulator developed by Mick West[15]). Proponents of this theory claim that pod roll must be avoided as much as possible because it requires moving the entire electro-optical system, a significant mechanical constraint given its heavy weight. However, we note that in this "step-rotation" theory, the pod rolls just as much as if it was rotating smoothly (i.e., it still needs to roll by the same angle). But in this case this is coupled with the additional mechanical constraint of having to stop and reactivate the roll multiple times, while rapidly recentering the internal mirrors over a few degrees (multiple FOV in the NAR FOV). Neither the patent behind the ATFLIR gimbal-mounted camera,[16] nor two ATFLIR experts that we questioned through private communications, describe such intermittency in pod roll being a feature of the ATFLIR electro-optical system. Rather, they suggested that the different axes of motion of the ATFLIR operate in concert to ensure smooth tracking of the target. A mechanical failure may explain the step-wise roll movement, but a malfunctioning targeting pod is yet another oddity in the chain of events required for the "distant plane" scenario. We recognize this is well beyond our expertise, and encourage other experts to share their views on this aspect of the Gimbal video (within the limits imposed by the classified nature of such military systems).

[13] https://www.metabunk.org/sitrec/
[14] https://www.nytimes.com/2022/10/28/us/politics/ufo-military-reports.html

[15] https://www.metabunk.org/gimbal/
[16] https://patents.google.com/patent/US9121758B2/en

There still is a lively debate around this video, which could probably be settled if more experts comment on the plausibility of the various scenarios. Regardless of where the truth lies, we encourage interested readers to explore the potential flight paths for Gimbal using the Sitrec simulator. This is a comprehensive tool that is valuable to understanding the case and forming an opinion. The codes and data we have used for this study are also publicly available, so that different flight path scenarios can also be tested using our model.

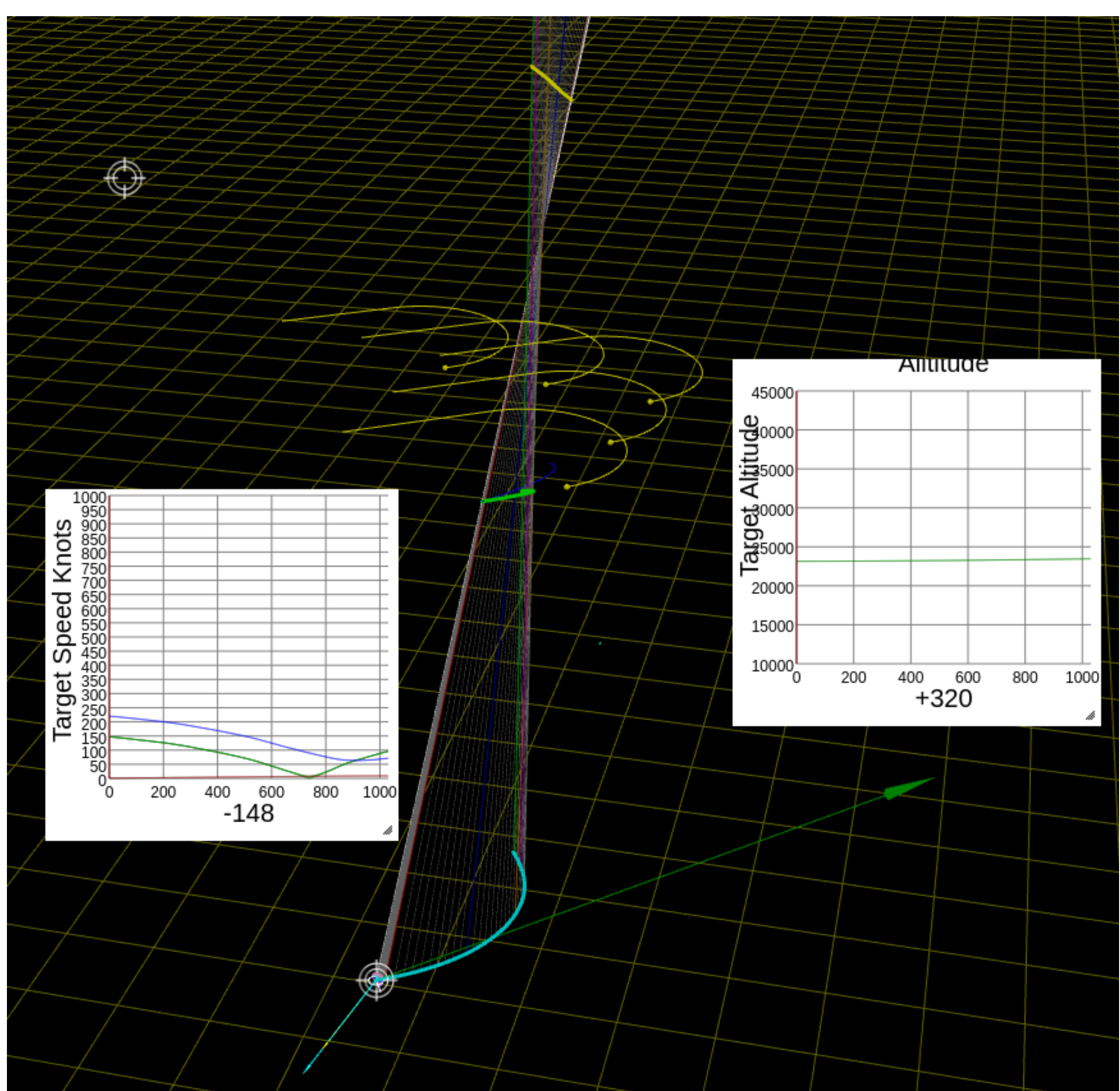

**Figure 10. Close flight path at 9 Nm (qualitatively similar to our flight path in Figure 6) in the Sitrec 3-D simulator (green curve). The blue curve shows the F/A-18F flight path, when facing a 120-kt wind (blue arrow for direction). The green arrow shows the F/A-18F initial heading. The yellow dots/curves represent the “fleet,” as mentioned in the cockpit audio. The distant yellow curve at ~30 Nm represents a potential flight path that would be consistent with a plane trajectory. The left graph shows the speed of the Gimbal object in this 9 Nm-scenario, with the green curve showing ground speed, and the blue curve air speed. The right graph shows the change in altitude of the object, climbing by ~300 ft along the flight path.**

### C. Discussion of other potential scenarios

*U.S. Technology*: One possible explanation for the Gimbal UAP's anomalous flight characteristics is that naval aviators unwittingly encountered an advanced U.S. test platform/prototype aircraft. However, to the best of our understanding, the aircrew were not required to sign nondisclosure agreements following the encounter (generally standard practice when military personnel are inadvertently exposed to classified technology). Moreover, in May 2022 testimony to the House Intelligence Committee, Deputy Director of Naval Intelligence Scott Bray and Undersecretary of Defense for Intelligence and Security Ronald Moultrie expressed high confidence that the first 144 reports analyzed by the UAP Task Force[17] did not include incidents involving secret or experimental U.S. aircraft. Moreover, operating or testing advanced, classified technology in tightly controlled training airspace amid a large-scale pre-combat deployment exercise raises a host of safety of flight and other operations security (OPSEC) issues, especially when the Department of Defense operates dedicated, secluded testing and evaluation ranges not over international waters.

*Foreign Surveillance/Drone*: A January 12, 2023, *New York Times* story, citing government officials, states that the Gimbal video (along with two other U.S. Navy UAP videos) has "not been categorized as [a foreign] surveillance [incident], at least so far."[18] Of note, at the beginning of the video, the F/A-18F pilot mentions the possibility that the object is a drone. Although this seems more plausible to us than a distant plane, the remote location, high altitude, small (15-20 ft) size of the object, lack of wings/propulsion, and sharp maneuver in the vertical require advanced technology that is not familiar to us. Here, again, comments from military/aeronautics experts would be beneficial, but such a brazen demonstration of highly advanced flight characteristics does not align with standard electronic intelligence tradecraft/OPSEC for breakthrough technology.

*V/STOL/Rotary Wing Aircraft*: To the best of the authors' understanding, the vast majority of V/STOL aircraft and rotary wing aircraft are incapable of engaging in hover/near-hover at ~23,000 ft (with conventional helicopters capable of conducting similar maneuvers at a maximum altitude of ~11,000 ft). At the same time, such aircraft are generally incapable of traveling at the 300-400 knot speeds retrieved within the lines of sight. Nor does the IR signature seem to fit a V/STOL/rotary wing aircraft.

*Electronic Warfare/Radar Error*: Some analysts suggest that the radar tracks observed by the Gimbal aircrew may be glitches or electronic warfare. The USS Roosevelt strike group was in the middle of a COMPTUEX mission, so this has been considered by the aviators (Figure 1). In short, this theory requires that electronic warfare technology was tested on/deployed against unwitting aviators (not standard practice), or the inability of Navy personnel to recognize instrument failure (or their own air-to-air exercise) when reporting a UAP encounter. One theory supposes that the Gimbal object is the engine exhaust from a distant plane actively spoofing the F/A-18F's radar, reducing the radar-observed distance to the aircraft (<10 Nm instead of 30 Nm). In this scenario, the Gimbal UAP and the "fleet" of 4-6 objects accompanying it are false radar tracks injected electronically by the adversarial plane. However, a review of open-source data suggests that "link/synthetic inject-to-live" technology and/or P5 tactical combat training system-style pods, which populate simulated radar tracks for training purposes, were not in use on aircraft carrier-borne units in early 2015. Moreover, this theory supposes that electronic warfare was not disclosed to the squadron or identified during a post-mission debrief involving intelligence personnel and a flag officer. Nor does this theory adequately explain the Gimbal object's unusual IR signature, unless this was coupled with unique IR signature management. When asked, LT Ryan Graves stated that he does not believe that the Gimbal incident was part of an electronic warfare exercise. An alternative version of this theory requires that a hostile foreign power's aircraft was not identified by a carrier strike group (outboard radar data) while flying in an active Navy training range.

---

[17] https://www.dni.gov/files/ODNI/documents/assessments/Prelimary-Assessment-UAP-20210625.pdf
[18] https://www.nytimes.com/2022/10/28/us/politics/ufo-military-reports.html

## V. Conclusion

Three-dimensional geometric reconstructions of the Gimbal UAP event include solutions that match witness descriptions of a highly anomalous flight path. At the range provided by witnesses (i.e., within 10 nautical miles from the F/A-18F), geometrical reconstructions show that the Gimbal object must stop and reverse direction along the vertical, in what is best described as a "vertical U-turn." This explains the brief stop and the absence of radius of turn observed on the aircrew's situational awareness (SA) radar display. The vertical U-turn also coincides with the long, continuous rotation of the object observed in the FLIR video. Including the effect of wind is critical and helps explain why the stop on the SA page was observed just before the rotation in the infrared video. Of course, the solution we highlight here depends on the accuracy of the provided radar range (< 10 Nm), which is not available to us. Although we depend on (high confidence) witness testimony for that aspect of the data, the match between the reconstructed trajectories, witness accounts, and what is observed in the video is remarkable.

Although uncertainties in several parameters limit a precise assessment of the speed and maneuvers of the object, our results support that the Gimbal object exhibited anomalous flight characteristics. An ability to maintain low speed at high altitude, without apparent (large) wings to compensate for low air density at altitude, and the ability to reverse direction in the vertical are two perplexing features. For example, a fighter jet requires several thousand feet to conduct such a direction reversal in the vertical. The Gimbal object, however, executes this maneuver in only a few hundred feet. The absence of an exhaust plume or other tell-tale signs of propellant/powered flight in the direction of motion raises additional questions about the nature of the object. Overall, this makes identification of the object difficult. An advanced drone may have the ability to make sharp maneuvers, but the range of velocities, lack of wings (especially in conjunction with the high altitude), remote location, and odd IR signature raise doubts about the plausibility of this explanation. At the same time, in the 0.35° field of view and at the 8 Nm range provided by witnesses, the object's infrared signature is approximately 15-20 feet along its longest axis. This suggests that the Gimbal object is markedly smaller than any conventional aircraft. Ultimately, however, the purpose of this study is to point out potential flight paths for the Gimbal object that align with witness accounts. Identifying the object is beyond our expertise.

While it has been suggested that the Gimbal UAP is simply a case of sensor-induced optical illusion and aircrew error, it is our opinion that this hypothesis does not fit the data. Even if we consider that the apparent rotation in the Gimbal video is caused by the roll of the ATFLIR pod, the context plus agreement between 3-D reconstructions and witness accounts suggest there is much more to this event than simply a distant plane locked by error, coupled with an unusual "glare" in the optics. Clearly, there is a link between what was observed on radar and the object in the FLIR, given how the reconstructed close paths align with what the aircrew saw on the situational awareness (radar) display. This means that if a distant plane was involved, the radar data was either erroneous or tampered with; both scenarios that are far beyond a simple and mundane misidentification by the aircrew. In sum, our opinion is that the evidence available to us points to a more straightforward scenario requiring fewer low-probability assumptions (i.e., the object was within 10 Nm of the F/A-18F and followed an anomalous flight path).

Through this work, we seek a more organized and transparent effort to investigate this case, to include experts from the aeronautics, engineering, and defense sectors. Moreover, if the Department of Defense still retains it, the public disclosure of radar data from this event would be extremely beneficial. If not in original format (e.g., due to classification restrictions), an official Department of Defense communication of the object's approximate range to the F/A-18F would suffice. As we have shown in this work, range is critical to refining potential flight paths. Dissemination of such information would go a long way in gaining a better understanding of the Gimbal incident. Of note, it is plausible that the F/A-18F WSO and/or pilot involved in the event may ultimately speak publicly. Such commentary would likely assist greatly in evaluating the plausibility of different scenarios. We look forward to any feedback these individuals may have on this study and will update the results with relevant information that may be released in the future.

Gimbal is a fascinating case, especially since it occurred amid daily, years-long observations of anomalous objects in tightly controlled training ranges [1]. Accurately characterizing the UAP observed by naval aviators off the U.S. East Coast is a matter of national security, aviation safety, and, based on the findings of this paper, possibly novel scientific knowledge. By raising awareness of the Gimbal incident in the aerospace and broader technical communities, it is our hope that this study will catalyze further interest and engagement in this case, and UAP more broadly.

## Appendix

**Size estimate for the object, based on a FOV of 0.35° for the ATFLIR**

$$size(ft) \; = \; 2 \times tan(FOV/2) \times distance\;(Nm) \times \; 6076.12\;(Nm\;to\;ft) \times fraction\;FOV$$

For an initial distance of 8 Nm and considering that at the beginning the object spans between 1/15th and 1/20th of the horizontal FOV, this gives an estimated size for the object of 15-20 ft.

**Figures**

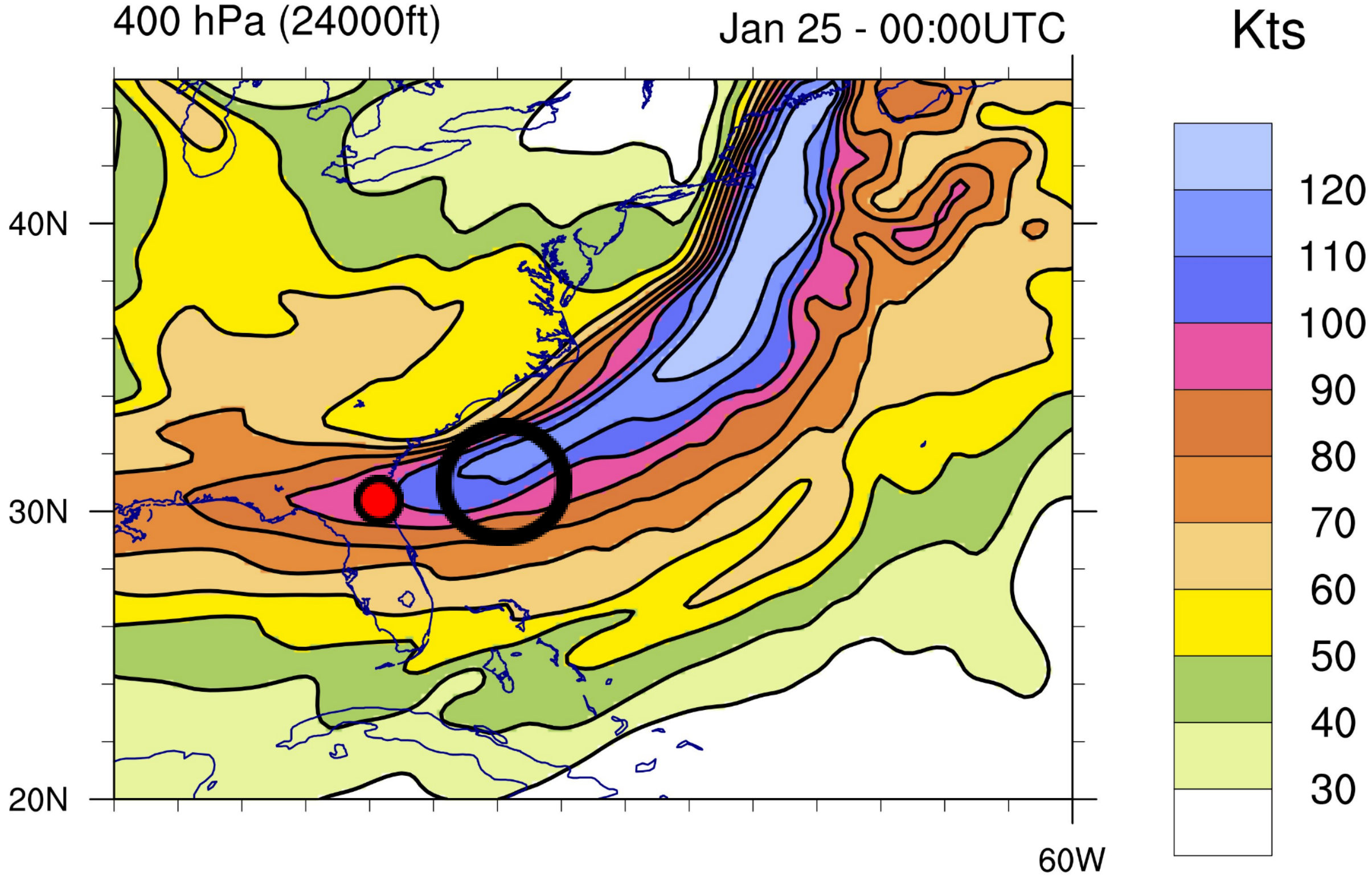

**Figure A1. Wind speed (knots) at 400 hPa (~24,000 ft at the latitude of Jacksonville, Florida) on January 25, 00:00 UTC (7 pm local EST). From the hourly ERA5 reanalysis. The black circle shows the approximate location for the event (300 miles off the coast of Jacksonville, red dot).**

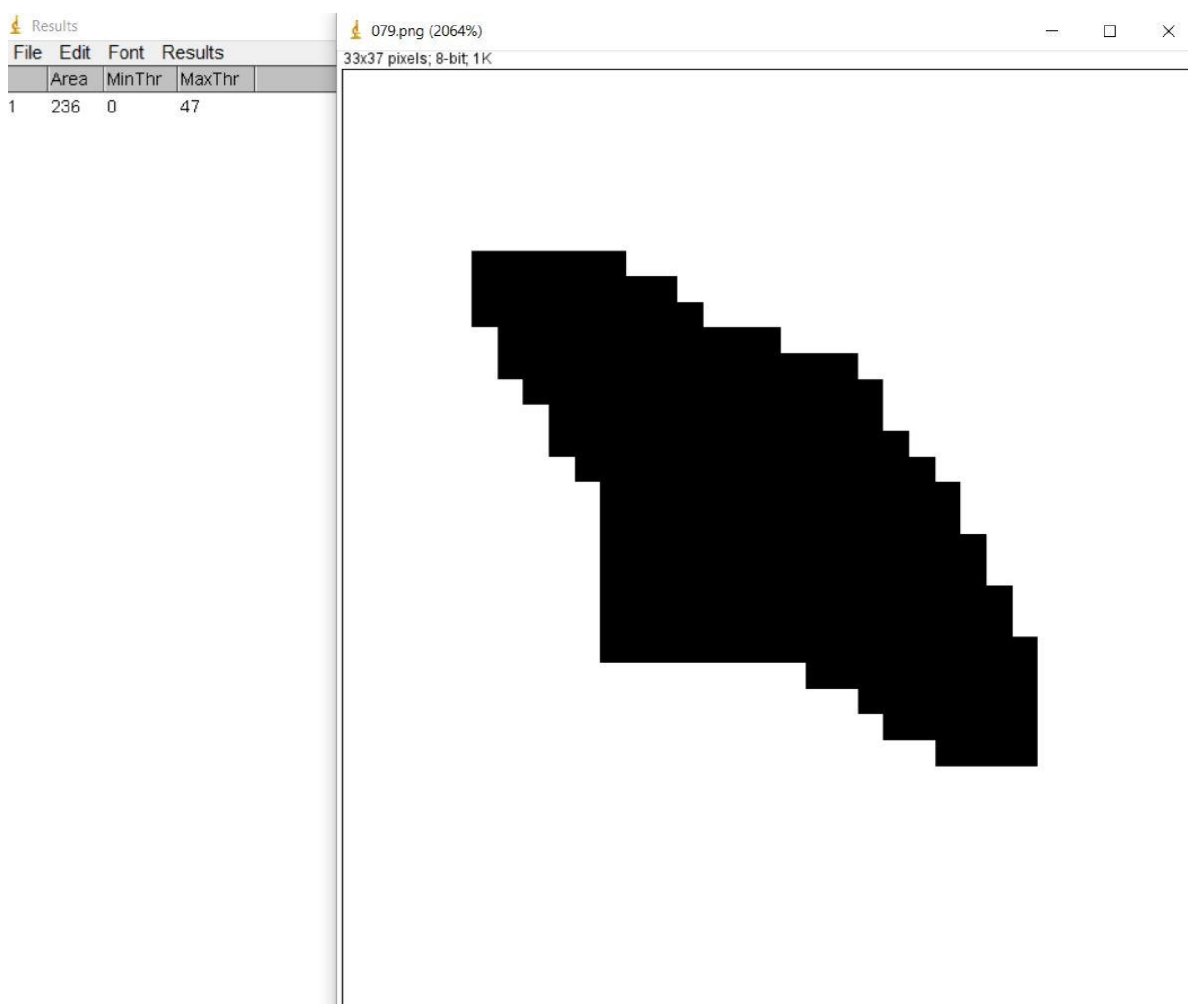

**Figure A2. Illustration of the method used to estimate the change in apparent size of the object. This was made using the ImageJ software [4] to count pixels above a constant grayscale threshold throughout 11 selected images (every 100 frames). The jump in apparent size from "white-hot" (WHT) to "black-hot" (BHT) mode was accounted for, to scale size estimates accordingly between the two modes. The number of pixels identified through this method increases by ~15% throughout the video.**

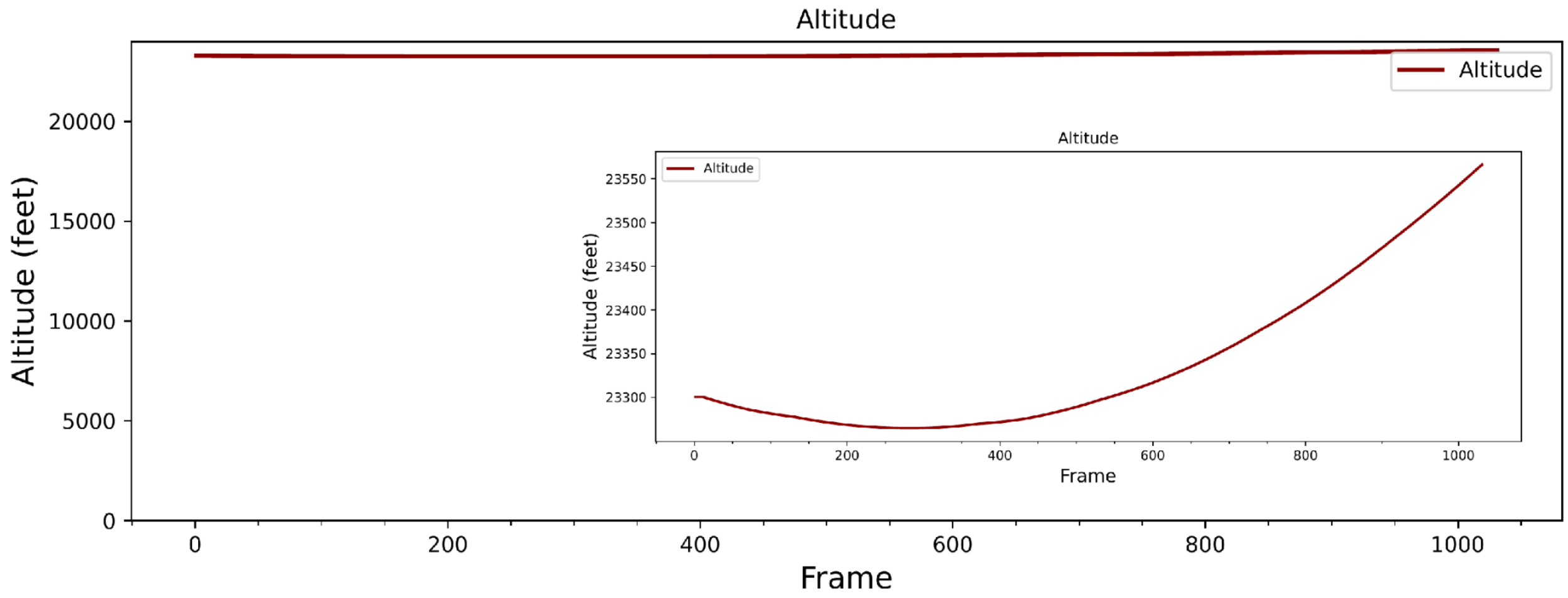

**Figure A3. Change in altitude of the object (feet), in the scenario shown in Figure 6. The object climbs by about 250 ft in this scenario (see refined values in the embedded plot).**

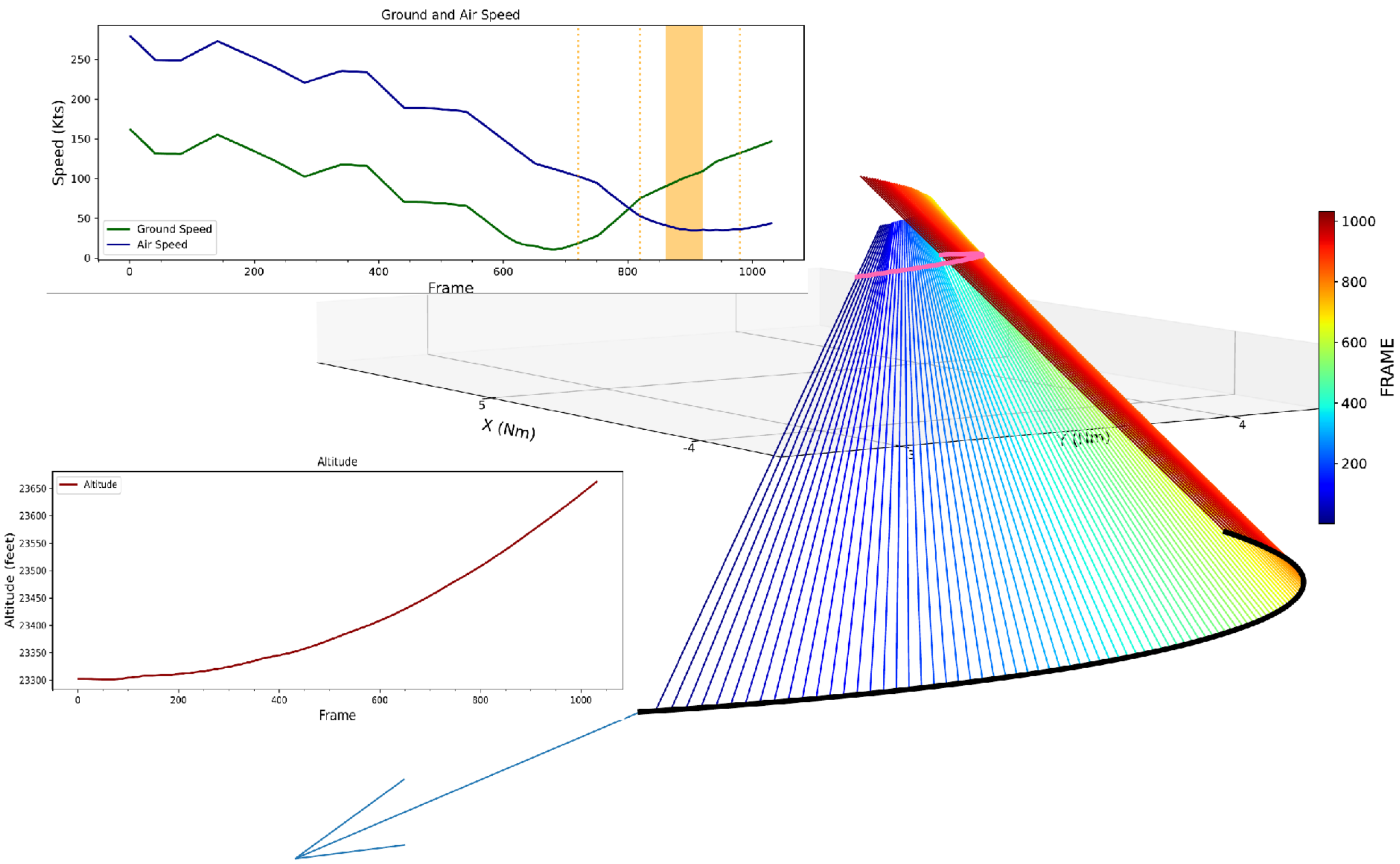

**Figure A4. Close flight path at 8 Nm (same wind/offset as in Figure 6), but using a field of view of 0.7°. This only impacts the adjustment of the lines of sight using background cloud motion. At close range this choice of FOV does not strongly affect the results.**

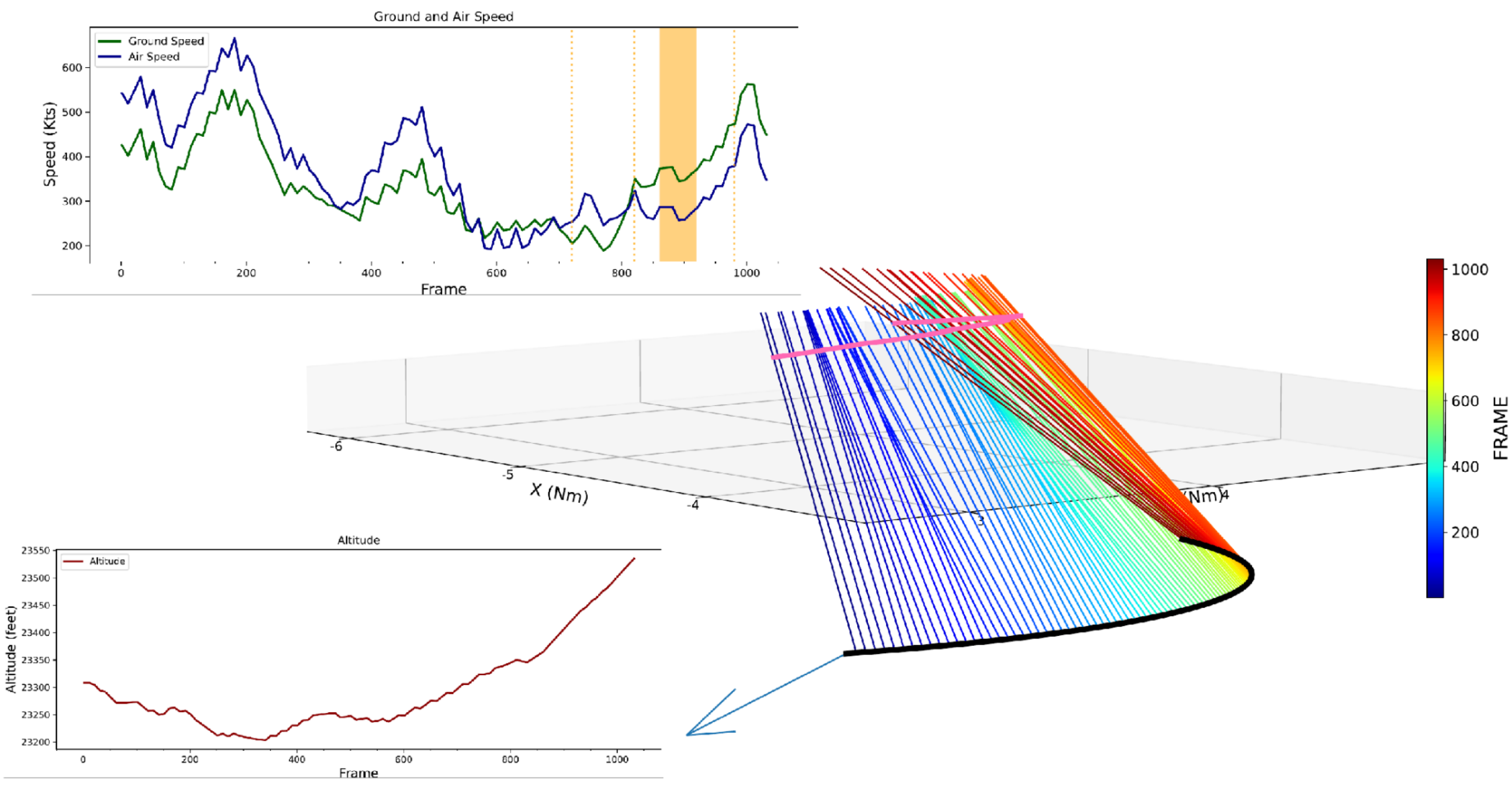

**Figure A5. Close flight path at 8 Nm (same wind/offset as in Figure 6), with the same FOV of 0.35°, but with no adjustment of the lines of sight using cloud motion.**

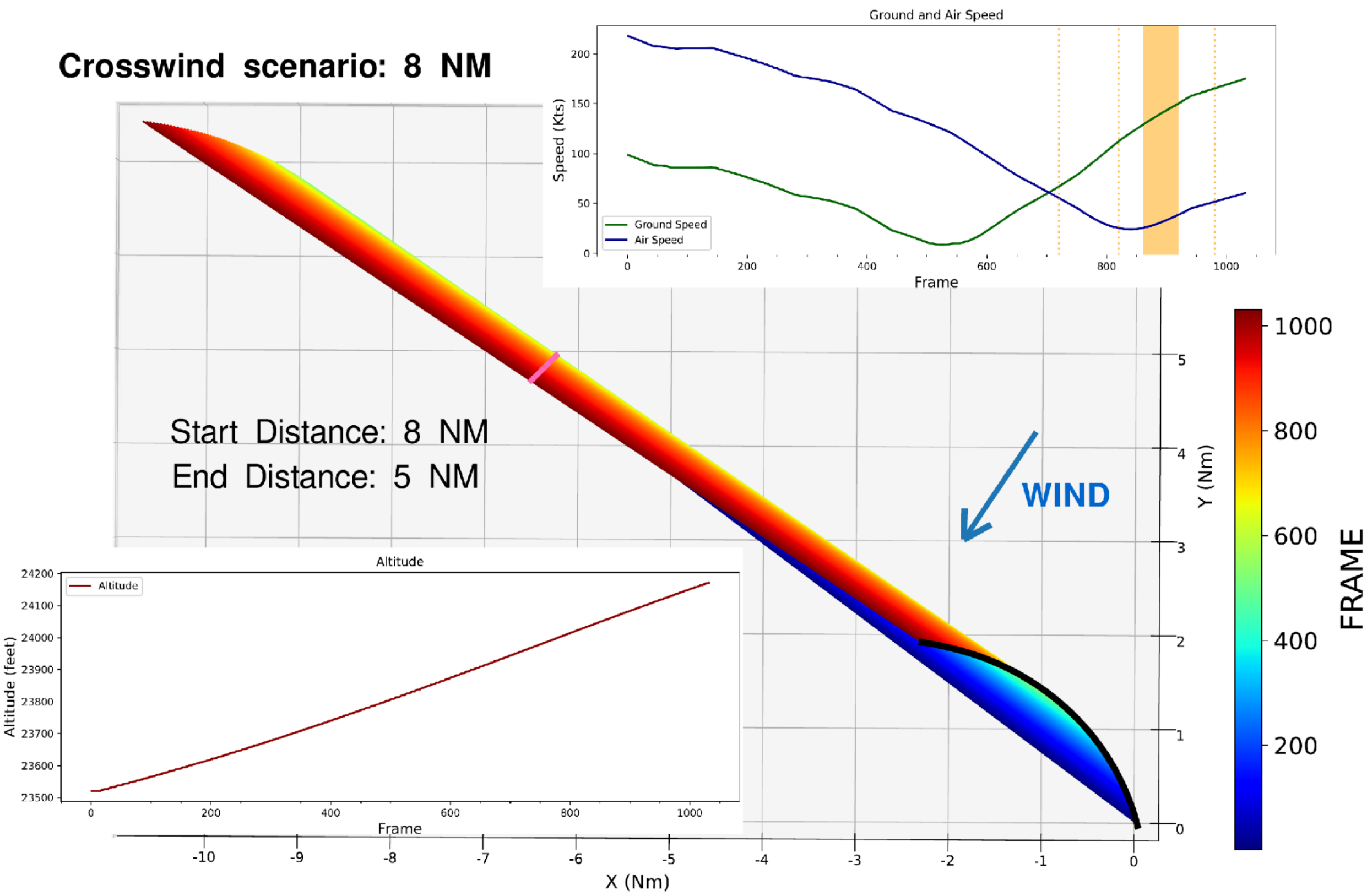

**Figure A6. Close flight path at 8 Nm with a wind direction oriented 35° to the right of the F/A-18F initial heading, and an offset of 10° relative to the wind for the object. This scenario implies slower speed than our scenario of Fig. 6, a greater altitude climb (~600ft) and a reduction of the distance to the object of 37%. This should correspond to an increase in apparent size of ~50%, not observed, which is why we favor the scenario in Figure 6.**

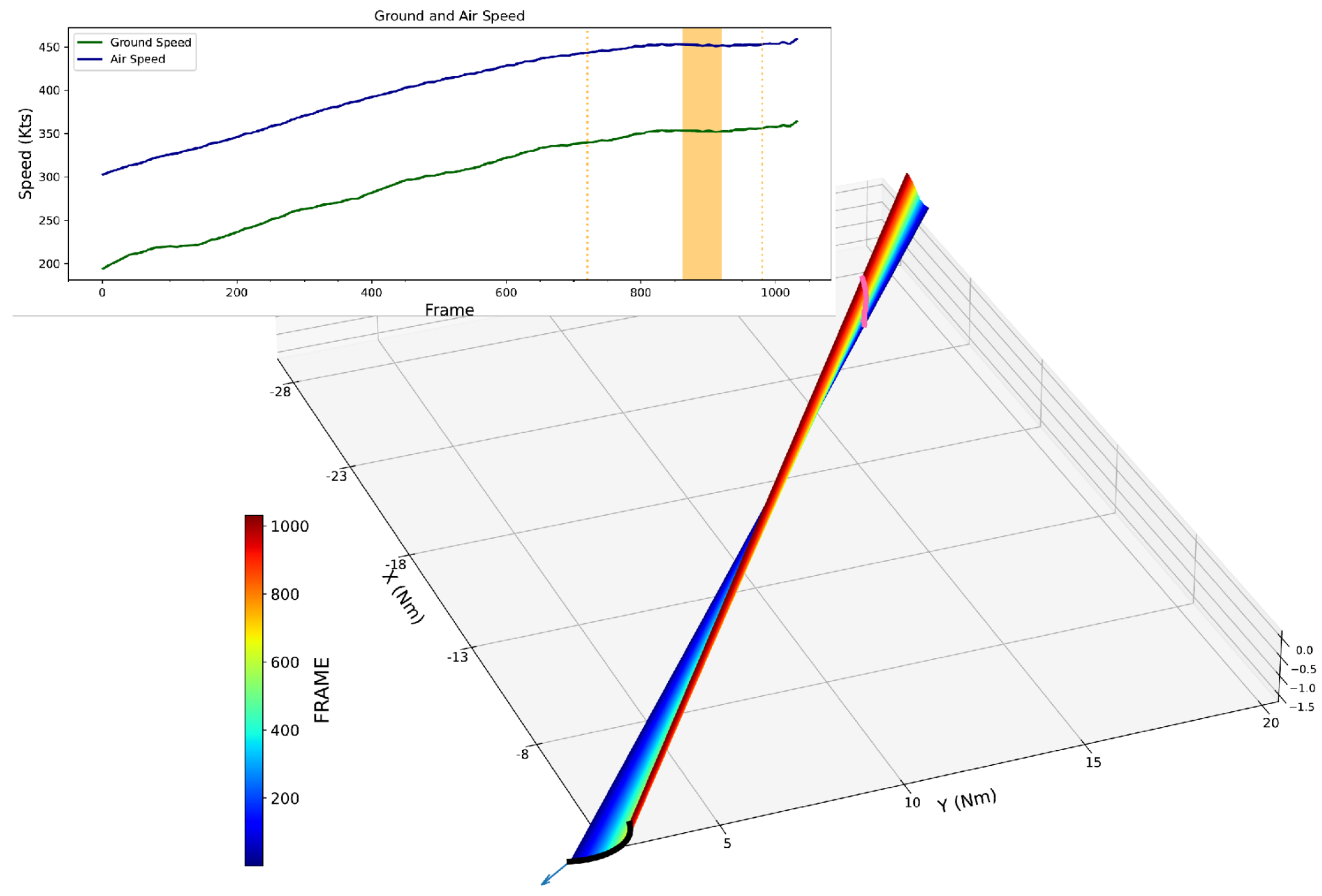

**Figure A7. Potential flight path at about 30 Nm from the F/A-18F, where fairly straight and leveled flight path solutions exist. The altitude is set constant to 18,700 ft to simulate a leveled flight path at 30 Nm. This is very consistent with the Sitrec simulator. In this configuration, however, we do not find a steady speed for the object (i.e., it needs to accelerate from 300 to 450 kts).**

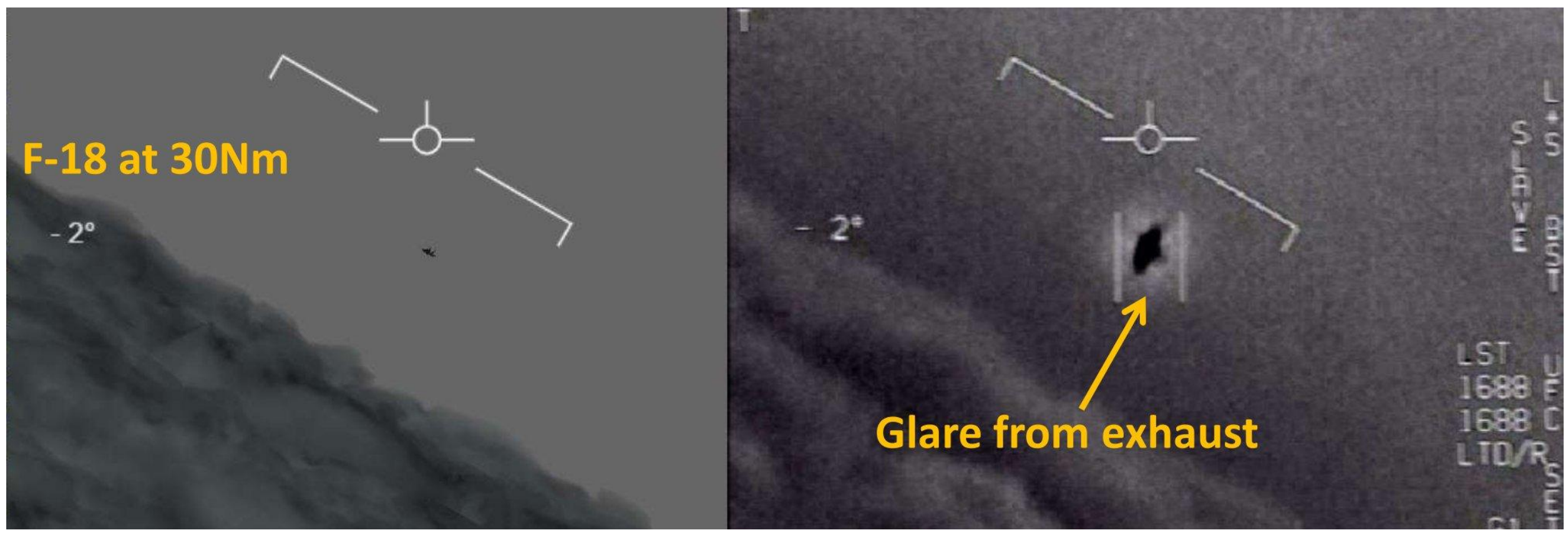

**Figure A8. Comparison between the estimated size of an F/A-18F at 30 nautical miles (left, in the 0.35° NAR2 FOV for the ATFLIR, https://www.metabunk.org/sitrec) and the end section of the Gimbal video (right). The NAR2 FOV is a digital zoom (x2) of the narrow optical zoom NAR1, FOV 0.7°. One pixel of the NAR1 FOV corresponds to ~4-5 ft at 30 Nm, following the formula: size(ft)=2*tan(FOV/2)*dist(Nm)*6076.12(Nm to ft)/480(# pixels on width). At that distance, the heat source (F/A-18F exhaust nozzles) has an equivalent size of just a few pixels of the infrared camera.**

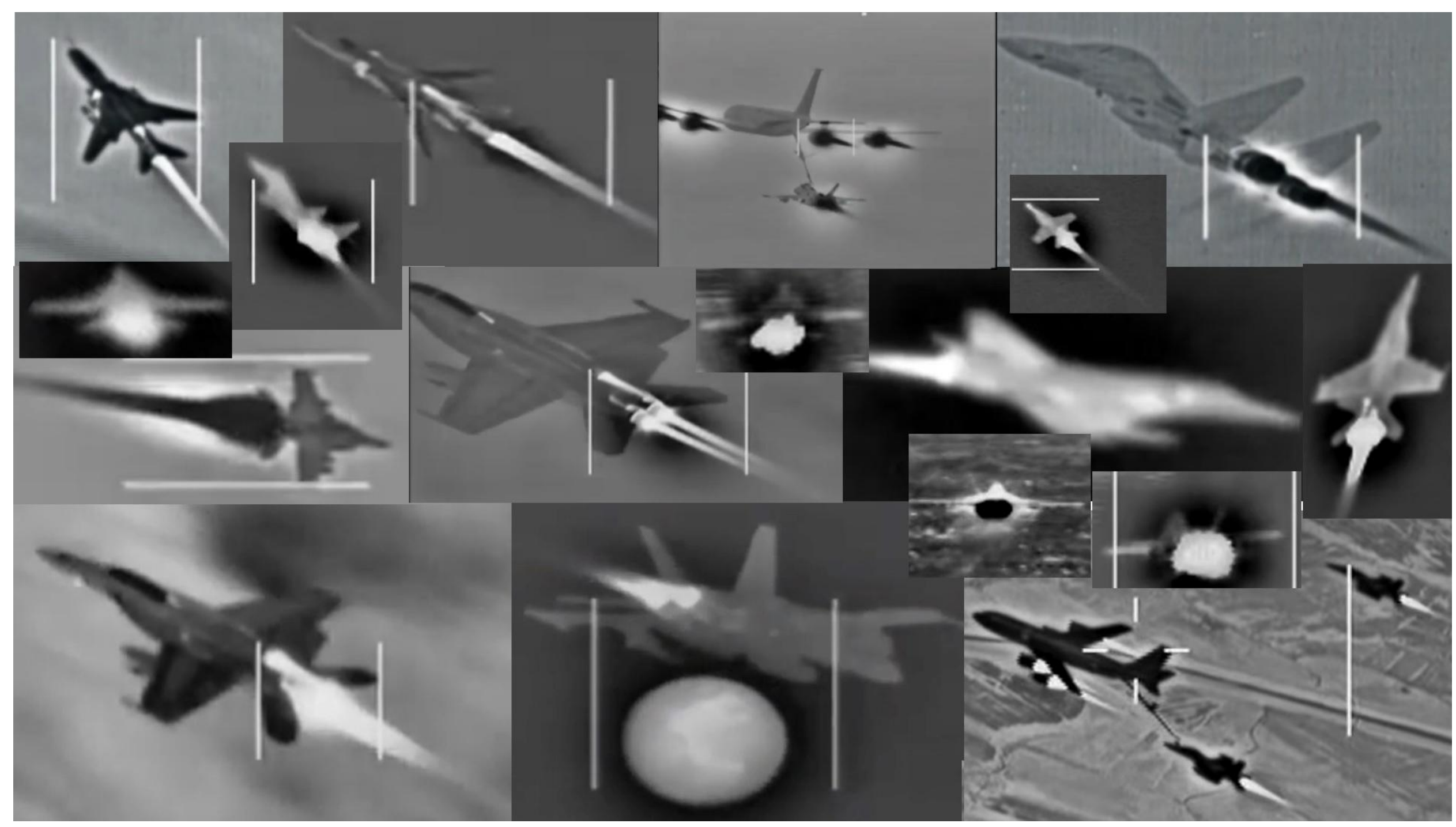

**Figure A9. Publicly available AN/ASQ-228 ATFLIR imagery of jet engine exhaust.**

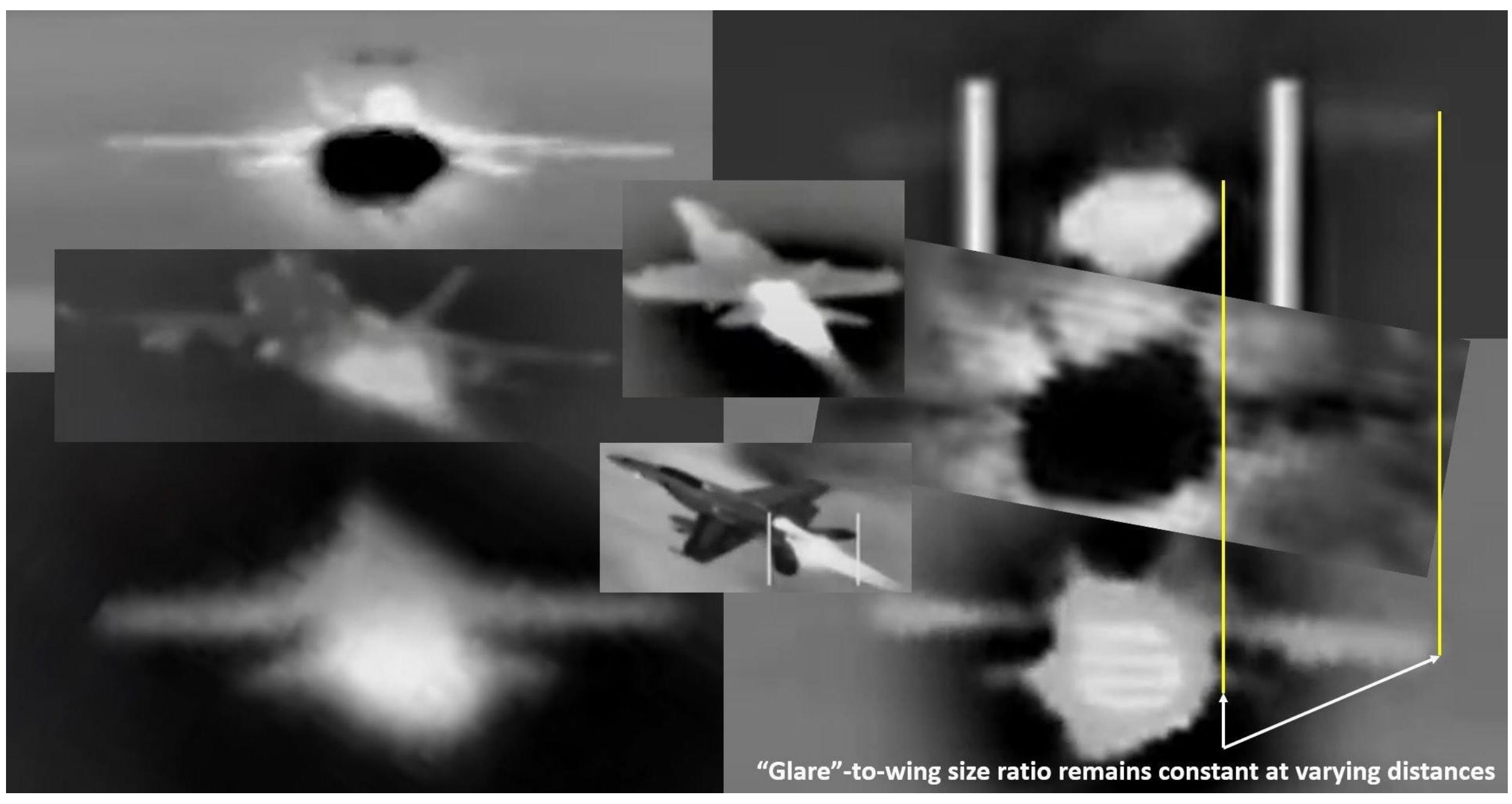

**Figure A10. Publicly available AN/ASQ-228 ATFLIR imagery of F/A-18 jet engine exhaust. In the available ATFLIR footage, exhaust viewed approximately tail-on never obscures the entire aircraft by several factors, as required by the "distant jet" theory (Figure A8).**

## Acknowledgments

The authors are grateful to LT Ryan Graves for providing us with his recollection of the event, including descriptions of the context and the SA page as observed during debrief on the USS Roosevelt. We also acknowledge the numerous private citizens who have contributed to efforts in understanding this case, on Twitter and on the Metabunk forum (user Edward Current being the first to point out the close-range vertical U-turn trajectories in the lines of sight, as well as potential solutions for a distant plane around 30 Nm). We are particularly grateful to Twitter users Zaine Michael, Chris Spitzer, Scott Manning, Spectre, Ray Moore, and #UFOTwitter in general. Special thanks to Mick West for providing the Sitrec simulator to the community. Thanks to The Black Vault (www.theblackvault.com) for information on Gimbal and other UAP cases obtained through the Freedom of Information Act. We finally thank Patrick Donovan, Paul Oberlin and the UAP AIAA committee for the opportunity to present this work in their session, and their help in refining the manuscript.

## Data availability statement

The Gimbal video is available on the U.S. Department of Defense website: https://www.defense.gov/News/Releases/Release/Article/2165713/statement-by-the-department-of-defense-on-the-release-of-historical-navy-videos/. The ERA5 reanalysis used for wind data is available at https://climate.copernicus.eu/climate-reanalysis. The source code and data that was used to produce the analyses and figures from this paper is publicly available at https://github.com/ypeings/Gimbal-UAP.git.